\documentclass[12pt]{iopart}
\usepackage{graphicx,iopams}
\usepackage{pstricks-add}
\usepackage{cite}

\usepackage[none]{hyphenat}

\def\pmb#1{\setbox0=\hbox{#1}
\kern-.025em\copy0\kern-\wd0 \kern.05em\copy0\kern-\wd0
\kern-.025em\raise.0433em\box0}

\newcommand{\text}[1]{\rm #1}

\begin{document}

\title[Partition function zeros for the Ising  model on networks]{Partition function zeros for the
Ising  model  on complete graphs and on annealed scale-free networks}

\author{M. Krasnytska$^{1,2,3}$, B. Berche$^{2,3}$, Yu. Holovatch$^{1,3}$, R. Kenna$^{3,4}$}

\address{$^{1}$
Institute for Condensed Matter Physics, National Acad.
Sci. of Ukraine, UA--79011 Lviv, Ukraine}

\address{$^{2}$
Institut Jean Lamour, CNRS/UMR 7198, Groupe de Physique Statistique,
Universit\'e de Lorraine, BP 70239, F-54506 Vand\oe uvre-les-Nancy
Cedex, France}

\address{$^{3}$
Doctoral College for the Statistical Physics of Complex Systems,
Leipzig-Lorraine-Lviv-Coventry $({\mathbb L}^4)$}

\address{$^{4}$
Applied Mathematics
Research Centre, Coventry University, Coventry CV1 5FB, United
Kingdom}

\date{Coventry, January 23, 2015}

\begin{abstract}
We analyze the partition function of the Ising model on graphs of
two different types: complete graphs, wherein all nodes are mutually
linked and annealed scale-free networks for which the degree
distribution decays as $P(k)\sim k^{-\lambda}$. We are interested in
zeros of the partition function in the cases of complex temperature
or complex external field (Fisher and Lee-Yang zeros respectively).
For the model on an annealed scale-free network, we find an integral
representation for the partition function which, in the case
$\lambda > 5$, reproduces the zeros for the Ising model on a
complete graph. For $3<\lambda < 5$ we derive the
$\lambda$-dependent angle at which the Fisher zeros impact onto the
real temperature axis. This, in turn, gives access to the
$\lambda$-dependent universal values of the critical exponents and
critical amplitudes ratios. Our analysis of the Lee-Yang zeros
reveals a difference in their behaviour for the Ising model on a
complete graph and on an annealed scale-free network when $3<\lambda
<5$. Whereas in the former case the zeros are purely imaginary, they
have a non zero real part in latter case, so that the celebrated
Lee-Yang circle theorem is violated.
\end{abstract}

\submitto{\JPA}

\eads{\mailto{kras$_-$mariana@icmp.lviv.ua},
\mailto{bertrand.berche@univ-lorraine.fr},
\mailto{hol@icmp.lviv.ua}, \mailto{R.Kenna@coventry.ac.uk}}

\maketitle


\vskip 1cm

\section{Introduction}\label{0}

Since the pioneering works of Lee and Yang \cite{LeeYang52}, as well
as Fisher \cite{Fisher65}, analysis of partition function zeros in
the complex plane has become a standard tool to study properties of
phase transitions in various systems \cite{Wu08}, lattice spin
models being one of them.

The Lee-Yang zeros are calculated at (real) temperature ($T$) in the
complex magnetic field ($H$) plane whereas Fisher zeros (usually
studied in the absence of a magnetic field) are located in the
complex temperature plane. In the thermodynamic limit when the
system size $N$ approaches infinity, the Lee-Yang and Fisher zeros
form curves on the complex ($H$ or $T$) plane. By analysing the
location and scaling of these  zeros, an alternative description of
critical phenomena is achieved involving angles formed by these
curves \cite{Pearson82,Itzykson83,Inzykson85,Abe67,JaJoKe06}. In
this way the angles may be considered to be conjugate to the set of
critical exponents and critical amplitudes ratios. Of special
interest are the finite-size scaling properties of zeros, which also
encode universal features of
 underlying phase transitions \cite{Itzykson83}.

Besides being of fundamental interest, and being useful for theory,
the zeroes attract attention due to their experimental observation
too. The first step to connect zeros to experimental data was made
in Ref.\cite{Binek}. The density  of zeros on the Lee-Yang circle
was determined by analyzing isothermal magnetization data of the
Ising ferromagnets. Recently Peng et al. pointed out to possible
relation between imaginary magnetic fields and quantum coherence of
probe spin in a bath \cite{Wei12,Peng15}. In turn, the
experimentally obtained times at which quantum coherence disappears
are equivalent to imaginary Lee-Yang zeros. Such demonstration of
experimental realization of Lee-Yang zeros is important at a
fundamental level and points to new ways of studying zeros in
complex, many-bodied materials. E.g., the approach could help in the
study of real systems in which the zeros can't easily be calculated
giving access to new quantum phenomena that would otherwise remain
hidden if one were to restrict attention to real, physical
parameters. The experiments also confirm a profound connection
between a static entity (the complex magnetic field in
thermodynamics) and dynamical properties of quantum systems, namely
coherence.

The aim of our paper is to analyse  partition function zeros for a
spin model on a complex network \cite{networks}. {{In particular, we
are interested in a scale-free network with power-law decay of a
node degree distribution $P(K)\sim K^{-\lambda}$, where $P(K)$ is
the probability that network node has degree $K$. Such networks
attract much attention due to their applications in descriptions of
numerous natural and man-made systems, see e.g.
\cite{networks,Dorogovtsev08}.}} It is by now well established that
models on such networks have qualitatively similar critical behavior
to those on systems defined on regular $d$-dimensional lattices. In
particular, the exponent $\lambda$ determines collective behaviour
and   plays a role similar to that of the space dimension $d$ for
lattice systems; there are lower and upper critical values of
$\lambda$ for networks, analogous to lower and upper critical
dimensions for lattices. At the upper critical values $\lambda =
\lambda_{\rm{uc}}$ and $d= d_{\rm{uc}}$, scaling behaviour is
modified by multiplicative logarithmic corrections, while above
them, critical exponents assume their mean-field values. The above
analogy has limitations; e.g.,  for $d \leq d_{\rm{lc}}$ lattice
systems remain disordered at any finite temperature $T$ whereas
systems on networks are always ordered at $\lambda \leq
\lambda_{\rm{lc}}$. But the feature we want to emphasize by this
comparison and which will be central for the analysis presented
below is that $d$  and $\lambda$ play the role of {\em global}
variables determining universal properties of critical behaviour.
Therefore the parameter $\lambda$, like the dimensionality of a
regular lattice, governs the un\-iversality of thermodynamic
properties near the critical point along with the field symmetries
and range  of interactions. So, in the absence of a space
dimensionality, $\lambda$ controls the transition. For the Ising
model, as most models, $\lambda_{\rm{lc}} = 3$, $\lambda_{\rm{uc}} =
5$ while $d_{\rm{lc}} = 1$ and $d_{\rm{uc}}=4$.

Here we consider the so-called annealed network (see e.g. \cite{Lee09}).
For a spin model on such a network, the network configuration is
fluctuating together with spins, which leads to the partition
function configurational averaging \cite{Brout59}. In turn, this
allows one to attain an asymptotically exact description of many spin
models, placed on such a type of  network \cite{annealed}. In this
description, the partition function is represented in the form of the model
on a complete graph with random separable interactions \cite{Lee09}.
Therefore, in our analysis we
 start from the Ising model on a complete graph \cite{Glasser86,
Uzelac13}, considering properties of its partition function zeros in
the complex  $H$ and $T$ planes and then we will generalize this
analysis for the case of an annealed scale-free network. The
remainder of this paper is organised as follows. In
section~\ref{Sec-ComplexZeros} we briefly introduce the main
relations which follow from the properties of partition function
zeros and fix the notations. Section~\ref{II} delivers the analysis
of the Fisher and Lee-Yang zeros of the Ising model on a complete
graph. Following Ref. \cite{Glasser86} we represent the partition
function in  integral form, extracting the number of particles into
the variable of integration and complex reduced temperature or
magnetic field. In section ~\ref{III} we obtain the results for both
types of zeros for the Ising model on an annealed scale-free network
with a power-law decay of the node degree distribution characterised
by $\lambda$. {{We collect some details of our calculations in the
Appendices.}} Certain results of the sub-section \ref{IIIc} have
been previously announced in a Letter \cite{Krasnytska15}.

\section{Definitions and notations}\label{Sec-ComplexZeros}

In this section, we fix the notation and recall the derivation
of scaling relations required for the rest of the paper.
We wish to consider  the partition function either at the critical temperature $T=T_c$ in the complex magnetic field or at $H=0$ in complex temperature.
To investigate the former, we write $H={\rm Re}\,H+i\,{\rm Im}\,H$ and determine the Lee-Yang zeros $H=H_j$ of $Z(T_c,H)$ by searching numerically for the intersection in the complex
magnetic-field plane of the curves ${\rm Re}\,Z(T_c,H)=0$ and ${\rm
Im}\,Z(T_c,H)=0$.

\begin{figure}[t]
\centerline{\includegraphics[angle=0, width=8.5cm]{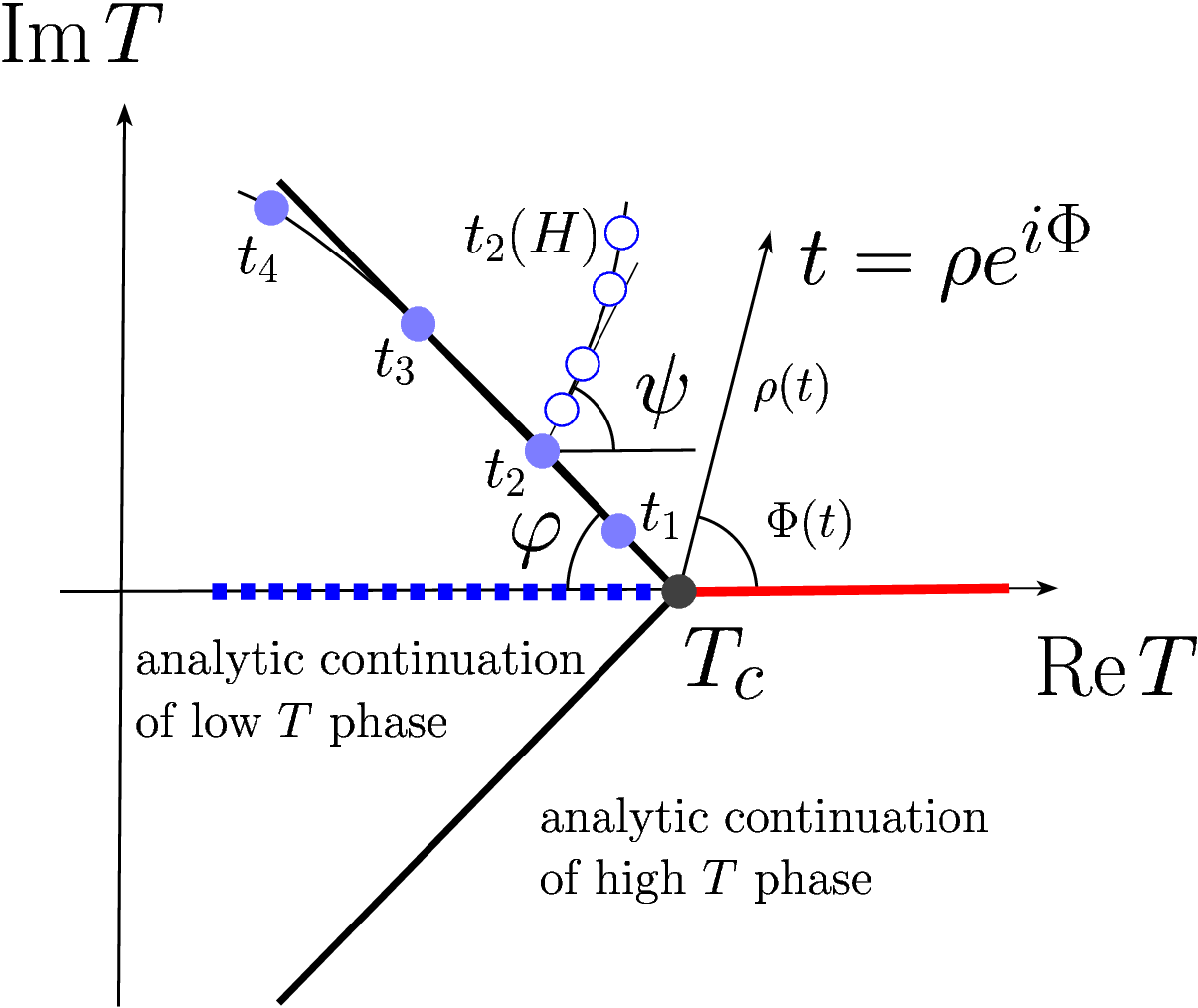}}
 \caption{
The distribution of Fisher zeros in the complex $T$ plane. The
variable $t = \rho e^{i\Phi}$ labels complex temperatures with
respect to $T_c$. The angle $\varphi$ is the impact angle of the
zeros with the negative sense of the real axis, so that $\Phi
\approx \pi - \varphi$ for the first few zeros which are indicated
by  discs. The angle $\psi$ describes the motion of the Fisher zeros
in presence of a real magnetic field. The motion of the first zero
with increasing $H$ is indicated by the circles.
  \label{fig1}}
\end{figure}

For our considerations in the complex temperature plane we write $T={\rm Re}\, T+i\, {\rm Im}\, T =T_c(1+t)$,
 where $T_c$ is the (real) critical temperature in zero field and $t=\rho e^{i\Phi}$ ($t=(T-T_c)/T_c$)
 parametrises  position  in the complex plane relative to $T_c$ \cite{Glasser87}.
 To determine the Fisher zeros, we search for intersections of ${\rm Re}\,Z(T,0)=0$ and ${\rm Im}\,Z(T,0)=0$.
These Fisher zeros accumulate in the vicinity of the critical point
$T_c$ along a line  and tend to pinch the positive real axis with
the angle $\Phi (t \to 0)= \varphi$ \cite{Pearson82, JaJoKe06,
Glasser87}. Consider the locus of  Fisher zeros in the complex
temperature plane depicted in Fig.~\ref{fig1}. Fisher zeros are
represented as light discs (blue online) the black one being the
critical point position. The critical temperature is real, so the
critical point is located on the real axis. Applying a real magnetic
field to the system, we observe that Fisher zeros move in the
complex $T$ plane along a curve which defines an angle $\psi$ with
the positive real axis~\cite{Itzykson83}.

A useful relation connects the impact angle $\varphi$ with the
exponent $\alpha$ of the specific heat and with the specific heat
universal amplitude ratio $A_-/A_+$, where $A_+$, $A_-$ are  scaling
amplitudes at $t>0$, $t<0$ respectively. On the real axis in the
thermodynamic limit, the singular part of the free energy in the
vicinity of critical point in the absence of magnetic field can be
written as $f\simeq F_{\pm}|t|^{2-\alpha}$. In the high-temperature
phase given by $t>0$, this is $f\simeq F_{+}t^{2-\alpha}$. In the
low-temperature phase for which $t<0$ it is  $f\simeq
F_{-}(-t)^{2-\alpha}$. In the complex plane, the free energy is
analytic everywhere except along the lines of zeros of the partition
function. That line separates the two regions (high and low
temperatures) in Fig.~\ref{fig1}. By analytic continuation from the
high temperature real axis,  the region on the right of the line of
zeros (the ``high temperature phase'') has
\begin{equation}\label{I.1}
f_+(t)\simeq F_{+}(\rho e^{i\Phi})^{2-\alpha}+f^{reg}_+,
\hspace{0.5cm} 0\leq\Phi< \pi-\varphi.
\end{equation}
On the other hand, continuing the region to the left of the line of zeros (the low temperature phase), one has
\begin{equation}\label{I.2}
f_-(t)\simeq F_{-}(-\rho e^{i\Phi})^{2-\alpha}+f^{reg}_-,
\hspace{0.5cm} \pi-\varphi<\Phi\leq\pi.
\end{equation}
At the transition along the line of Fisher zeros,
$\Phi=\pi-\varphi$, the real parts of the free energies of both
phases are equal \cite{Pearson82,Itzykson83}. Substituting
$\Phi=\pi-\varphi$ into (\ref{I.1})-(\ref{I.2}) and using the fact
that $F_-/F_+=A_-/A_+$  we arrive at the formula
\cite{Itzykson83,Abe67,JaJoKe06}:
 \begin{equation}\label{I.3}
\tan[{(2-\alpha)\varphi}]=\frac{\cos(\pi \alpha) -A_-/A_+}{\sin(\pi
\alpha)}.
\end{equation}

The scaling of the zeros follows from general arguments. Replacing
the volume $L^d$ of a regular lattice by the number of sites $N$ on
a graph the partition function in terms of complex reduced
temperature $t$ and magnetic field $H$, $Z(t,H)$, can be written as
a generalised homogenous function \cite{Itzykson83}:
\begin{equation}\label{I.4}
Z(t,H)=Z(tN^{1/(2-\alpha)},HN^{\beta\delta/(2-\alpha)}).
\end{equation}
The partition function is an  even function of $H$ so that the
previous scaling relation may be written either as
\begin{equation}\label{I.5}
H^2N^{2\beta\delta/(2-\alpha)}=f(tN^{1/(2-\alpha)}),
\end{equation}
or as
\begin{equation}\label{I.6}
tN^{1/(2-\alpha)}=g(HN^{\beta\delta/(2-\alpha)}).
\end{equation}
At $H=0$   equation (\ref{I.6}) gives the scaling of Fisher
zeros:
\begin{equation}\label{I.8}
t_j= N^{-1/(2-\alpha)} g_j(0),
\end{equation}
where $g_j(0)$ is in general a complex number. Extending to scaling
with the label index $j$~\cite{Itzykson83}  {we} generalize  from
(\ref{I.5}) and (\ref{I.6}) to the  Lee-Yang and Fisher zeros
scaling as
\begin{equation}\label{I.10}
H_j(N,t=0)\sim  \left(  \frac{j}{N}
\right)^{\frac{\beta\delta}{2-\alpha}}\, ,
\end{equation}
\begin{equation}\label{I.9}
t_j(N,H=0)\sim  \left(  \frac{j}{N}  \right)^{\frac{1}{2-\alpha}} \,
.
\end{equation}
Setting $t=0$ in (\ref{I.5}) leads for the Lee-Yang zeros to
\cite{Itzykson83}:
\begin{equation}\label{I.7}
 H^2_j=N^{-2\beta\delta/(2-\alpha)}f_j(0)\, .
\end{equation}
In general, $f_j(0)$ is a complex number. However, for models that
obey the Lee-Yang theorem  (all zeros are purely imaginary $H_j\sim
i\,{\rm Im}\, H_j$ \cite{LeeYang52}) $f_j(0)$ is a negative real
number. Using this property one can derive from the scaling
properties of the partition function \cite{Itzykson83} an angle
$\psi$ of motion of Fisher zeros  in real magnetic field. At fixed
$t$ and for $N\rightarrow \infty$, the variable $H_j$ tends to the
Lee-Yang edge, hence, substituting $N^{-1/(2-\alpha)}$ from
(\ref{I.8}) into (\ref{I.7}) we parametrise:
\begin{equation}\label{I.11}
t_j\sim H_j^{1/(\beta\delta)}\exp(\pm i \psi),
\end{equation}
to obtain \cite{Itzykson83}
\begin{equation}\label{I.12}
\psi=\frac{\pi}{2\beta\delta}.
\end{equation}

\section{Ising model on a complete graph}\label{II}

In this section we  consider the infinite-range version of the
Ising model which is equivalent to the mean-field version of its
lattice counterpart \cite{Kac,Stanley71}. It can be also regarded as
an Ising model on the complete graph, where each node of the graph is
connected to every other node making all interspin couplings equal
to each other. The Hamiltonian of the model reads
\begin{equation}\label{II.1}
-{\cal H}=\frac{1}{2N}\sum_{l\neq m}S_lS_m +H \, \sum _l S_l \, .
\end{equation}
Here, indices $l,\, m$ label the nodes of the graph so that
$(l,m)=1,\dots,N$; $S_l=\pm 1$ are the Ising spins; $H$ is an
external magnetic field and the sum $\sum_{l\neq m}$ spans all pairs
of nodes (not necessarily the nearest neighbours). The coupling
between the spins is taken to be inversely proportional to the
number of nodes, to make the model meaningful in the thermodynamic
limit $N\to \infty$. One can obtain an integral representation for
the partition function of the model (\ref{II.1}) by making use of
the equality $(\sum_{l}S_l)^2=N+\sum_{l\neq m}S_lS_m$ to write for
the $N$-particle partition function
\begin{equation}\label{II.2}
Z_N(T,H)={\rm Tr} \, e^{-\beta {\cal H}}=e^{-\frac{\beta}{2}}
\prod_{l=1}^{N}\sum_{S_l=\pm 1}\exp\Big(\frac{
\beta}{2N}(\sum_{l}S_l)^2+\beta H\sum _l S_l\Big)\, ,
  \end{equation}
and then applying Hubbard-Stratonovich transformation to express the
partition function in the form where summation over $S_l$ can be
taken exactly:
\begin{eqnarray}\label{II.3}
Z_N(T,H)= \\ \nonumber e^{-\frac{\beta}{2}} \sqrt{\frac{N\beta}{2
\pi}} \prod_{l=1}^{N}\sum_{S_l=\pm 1} \int_{-\infty}^{+\infty}
\exp\Big(\frac{-Nx^2\beta }{2}+\sum_{m}S_m \beta (x+H)\Big)dx \, .
\end{eqnarray}
Performing the summation, one arrives at the integral representation
for the partition function of the Ising model on the complete graph:
\begin{equation}\label{II.4}
Z_N (T,H)= \int_{-\infty}^{+\infty} \exp\Big (\frac{-Nx^2}{2T}+N\ln
\cosh[( x+H)/T]\Big)dx \, ,
  \end{equation}
where we have explicitly written the temperature $T=\beta^{-1}$
dependency taking the Boltzmann constant value $k_B=1$. In
(\ref{II.4}) and in all other partition function integral
representations below, we omit the irrelevant prefactors.

In classical settings, to get thermodynamic functions, the integral
(\ref{II.4}) is taken by the steepest descent method, see, e.g.
\cite{Stanley71}. In particular, the model undergoes a second order
phase transition at $T_c=1$ which is governed by the standard
mean-field values of the critical exponents:
\begin{equation}\label{II.5}
\alpha=0, \hspace{1em} \beta=1/2, \hspace{1em} \delta=3,
\hspace{1em} \gamma=1\, .
\end{equation}
For the sake of convenience it is appropriate to rewrite the
partition function (\ref{II.4}) in the reduced temperature
variable $t=(T-T_c)/T_c=T-1$. Changing the integration variable
$\sqrt{N}x/T\to x$ one gets \cite{Glasser86}:
\begin{equation}\label{II.6}
Z_N (t,H)= \int_{-\infty}^{+\infty} \exp\Big
(\frac{-x^2(t+1)}{2}+N\ln \cosh[ x/\sqrt{N}+H/(t+1)]\Big)dx\, ,\,
  \end{equation}
where the temperature dependent prefactor again has been omitted.
Being primarily interested in the properties of the partition
function itself, we  approximate (\ref{II.6}) by its expansion
at large $N$ and small $H$. Keeping the leading order contributions
in the linear in $H$ term, we expand the exponent function for
$N\rightarrow \infty$ and get \cite{Glasser86}:
\begin{equation}\label{II.7}
Z^{\rm exp}_N (t,H)= \int_{-\infty}^{+\infty} \exp\Big (\frac{-t\,
x^2}{2}-\frac{x^4}{12N} + \frac{x\sqrt{N}H}{t+1} +
O(1/N^{2})\Big)dx.
  \end{equation}

In the remainder of this section we  analyze the expressions for the
exact and approximated partition functions of the Ising model on a
complete graph, Eqs. (\ref{II.6}) and (\ref{II.7}).

\subsection{Fisher zeros for the Ising model on a complete graph at $H=0$}\label{IIa}

\begin{figure}[t]
\centerline{\includegraphics[angle=0,
width=8cm]{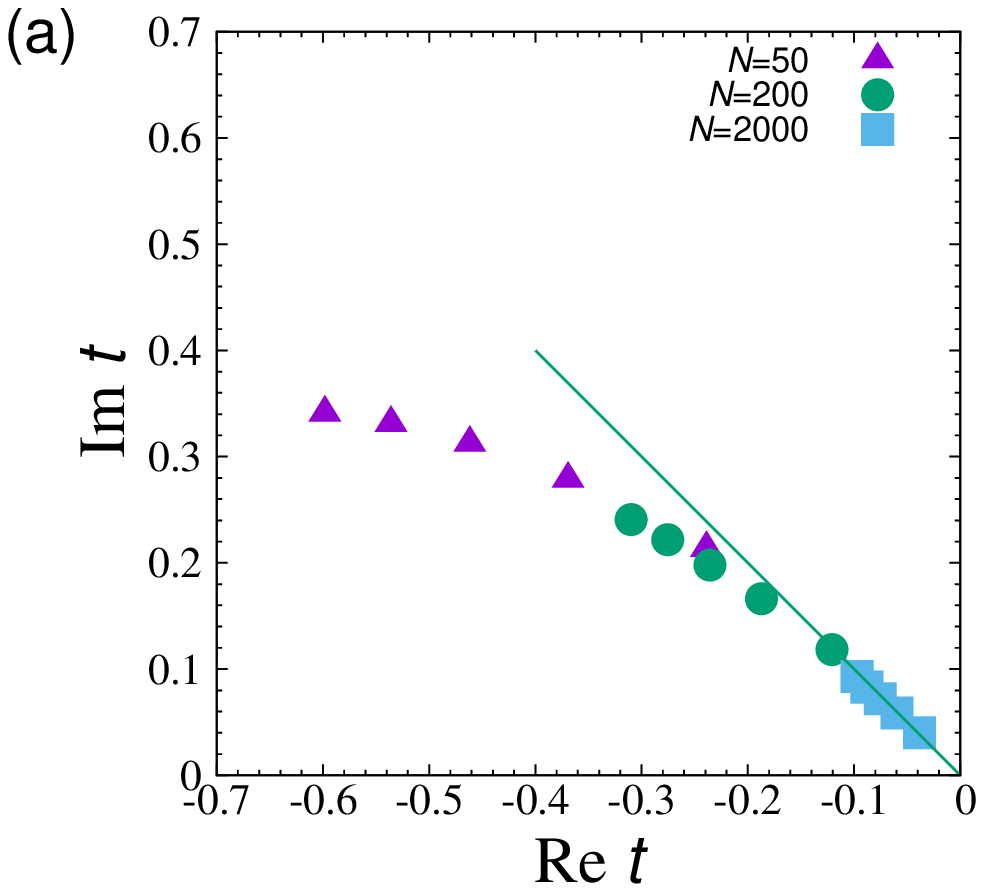}\includegraphics[angle=0,
width=8cm]{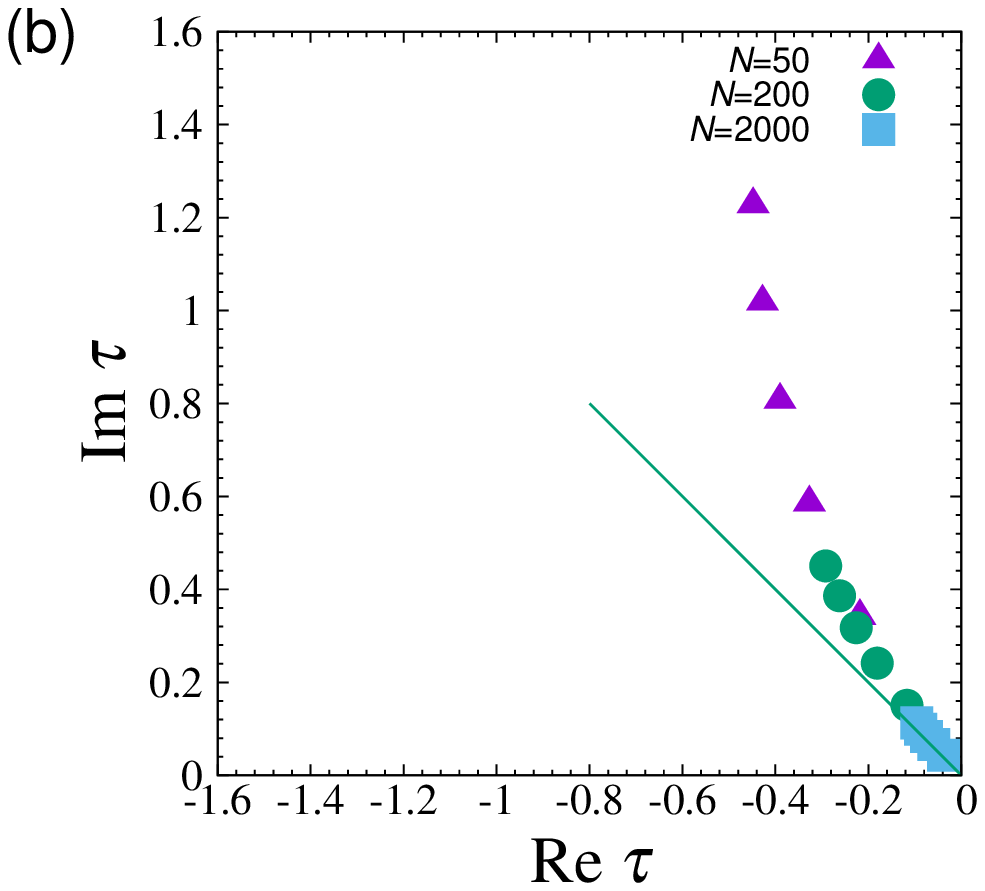}} \caption{The first five Fisher zeros for the
exact partition function (\ref{II.6}) with different values of $N$
(a) in the complex $t$ plane ($t=T-1$) and (b) in the complex $\tau$
plane $\tau=1-1/T$. With increasing $N$, the zeros accumulate at the
critical point, pinching the real axis at $t=\tau=0$. The solid line
has an angle $\varphi=\pi/4$ with the real axis. \label{fig2}}
\end{figure}

We start from the analysis of the exact integral representation for
the partition function at zero external field. First, we obtain the
Fisher zeros by solving the system of equations ${\rm Re}\,Z(t,H)=0$
and ${\rm Im}\,Z(t,H)=0$ for the complex variables $t={\rm Re}\, \,
t+ i\,{\rm Im}\, \, t$ using function (\ref{II.6}) at $H=0$. In
Fig.~\ref{fig2}{a} we show the first five Fisher zeros in the $t$
plane for  increasing values of $N=50, \, 200,\, 2000$. They
collapse on a master curve and in the vicinity of the critical
point, impact onto the real axis, defining the angle $\varphi$.

\begin{figure}[t]
\centerline{\includegraphics[angle=0, width=8cm]{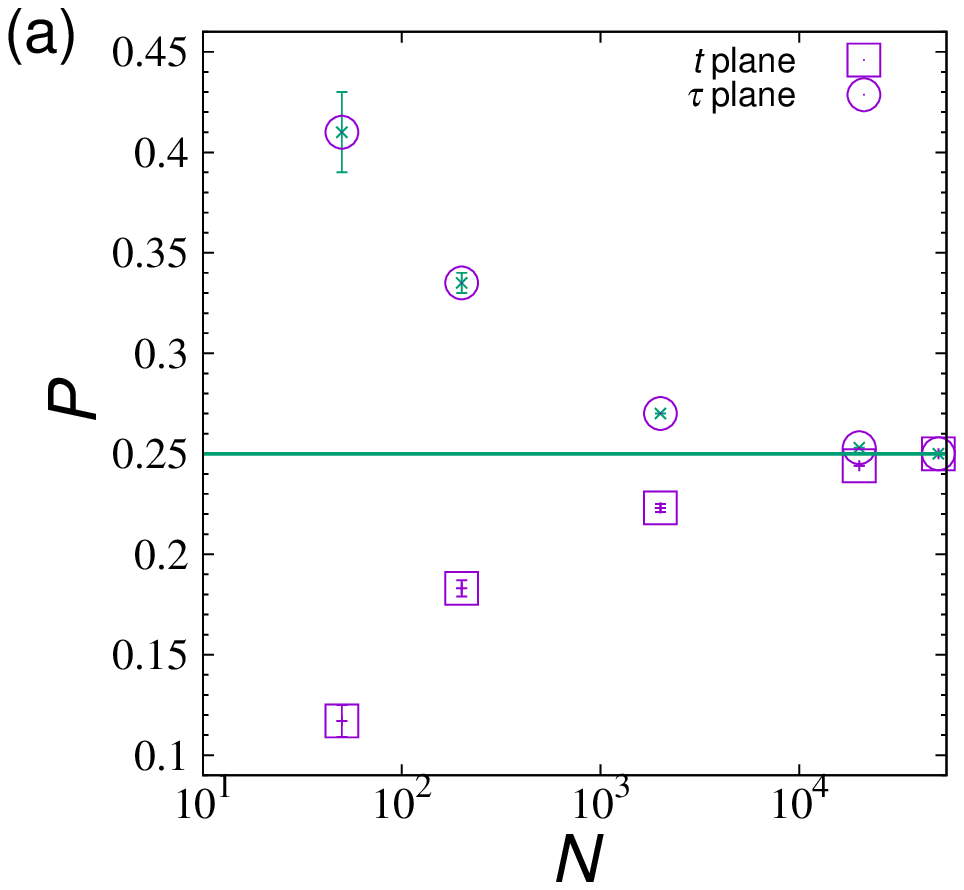}
\includegraphics[angle=0, width=8cm]{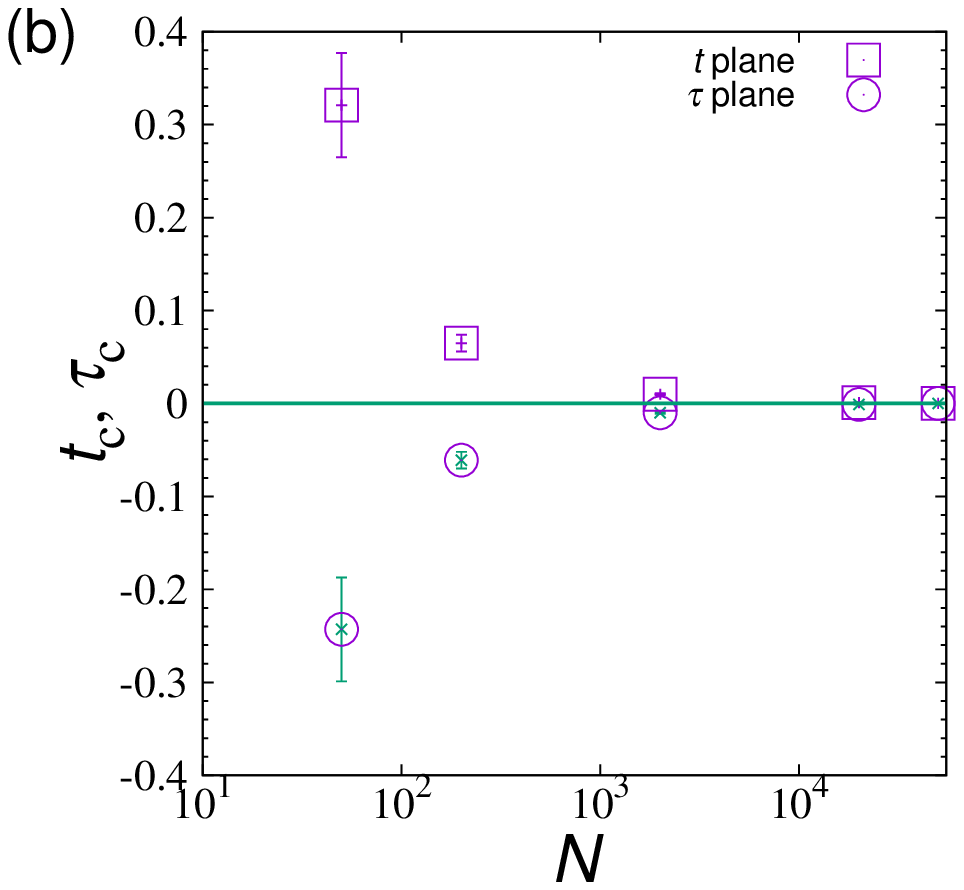}}
 \caption{Values of (a) the reduced impact angle $P= \varphi/
 \pi$
and of (b) the critical temperature for the exact partition function
obtained by fitting of the first five Fisher zeros for different
$N$. Squares: results in the complex $t$ plane; discs: results in
the complex $\tau$   plane; lines:  exact results. The larger the
system size, the more accurate is the fit. For $N>100$ the accuracy
interval is less than the size of the data point in the figure.
\label{fig3}}
\end{figure}

It is instructive also to observe the motion of the zeros if the
temperature is parameterised in a different way.  Let us introduce
the reduced inverse temperature:
\begin{equation}\label{II.8}
\tau=(1/T_c-1/T)/(1/T_c)=1-1/T\, ,
\end{equation}
getting from (\ref{II.4}) at $H=0$:
\begin{equation}\label{II.9}
Z_N(\tau)= \int_{-\infty}^{+\infty}
\exp\Big(\frac{-Nx^2(1-\tau)}{2}+N\ln \cosh [(1-\tau) x]\Big)dx.
\end{equation}
The zeros of the function (\ref{II.9}) in the complex $\tau$   plane
are shown in Fig.~\ref{fig2}{b}. As one can see from the figures,
the zeros form a smooth curve and  accumulate in the vicinity of the
critical point ($t=\tau=0$), and with an increase of $N$ they tend
to pinch the real positive temperature axis at the critical point.
Using known values of the critical exponents and amplitude ratios
one can evaluate the value of the angle $\varphi$ under which the
zeros pinch the real axis. Substituting (\ref{II.5}) into
(\ref{I.3}) and taking that the corresponding heat capacity
amplitude ratio $A_+/A_-=0$ one arrives at
\begin{equation}\label{II.10}
\varphi=\pi/4 \, .
\end{equation}
This value in shown in the Fig.~\ref{fig2} by the solid line. The
fact that the lines along which the zeros tend to accumulate with
increasing $N$ make the same angle $\varphi$ both in the $t$    and
$\tau$   planes illustrates the conformal invariance of this angle
-- angles are independent of any real analytic parametrization in
the complex plane. However, the approach of the angle to its value
in the thermodynamic limit $N\to \infty$ is parametrization
dependent. Indeed, depending on parametrization, the curves of
Fig.~\ref{fig2} reach the asymptotics from below ({{Panel~a}}) or
from above (Panel~b). This is further outlined in Fig.~\ref{fig3},
where we show how the values of the angle $\varphi$ and of the
critical temperature change with $N$.
To this end, we calculate the first five Fisher zeros for the system
sizes ranging up to $N=50000$, fit these points by a straight line
and find the values of the angle (taking $P=\varphi/ \pi$ so that
Eq.(\ref{II.10}) predicts $P=0.25$) and of the crossing point of
this line with the real axis. The larger the system size, the more
accurate the fit is (for $N>100$ the accuracy interval is less than
the size of the data point in the figure). Starting from $N=5000$,
$\varphi$ differs from its exact value ($\varphi=\pi/4$) by less
than 1\% while the critical temperature differs from the exact value
at $t=0$ by less than 1\% starting from $N=10000$.

Figs.~\ref{fig2} and~\ref{fig3} quantify the main features of the
behaviour of the Fisher zeros for the exact partition function
(\ref{II.4}) in the complex temperature plane. Results for the
impact angle $\varphi$ and for the critical temperature obtained on
their basis if compared with the exact results demonstrate two
tendencies: (i) they improve with an increase of $N$ and (ii) the
agreement is better for smaller values of the index $j$.

\begin{figure}[t]
\centerline{\includegraphics[angle=0, width=8cm]{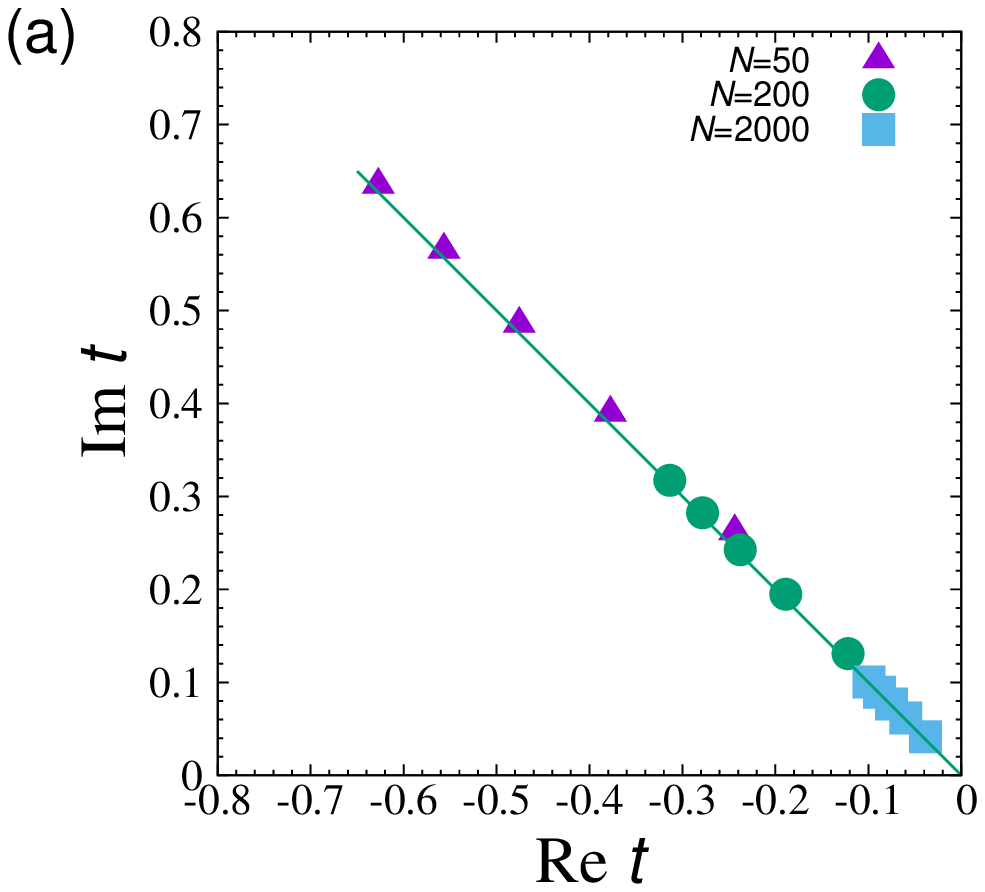}
\includegraphics[angle=0, width=8cm]{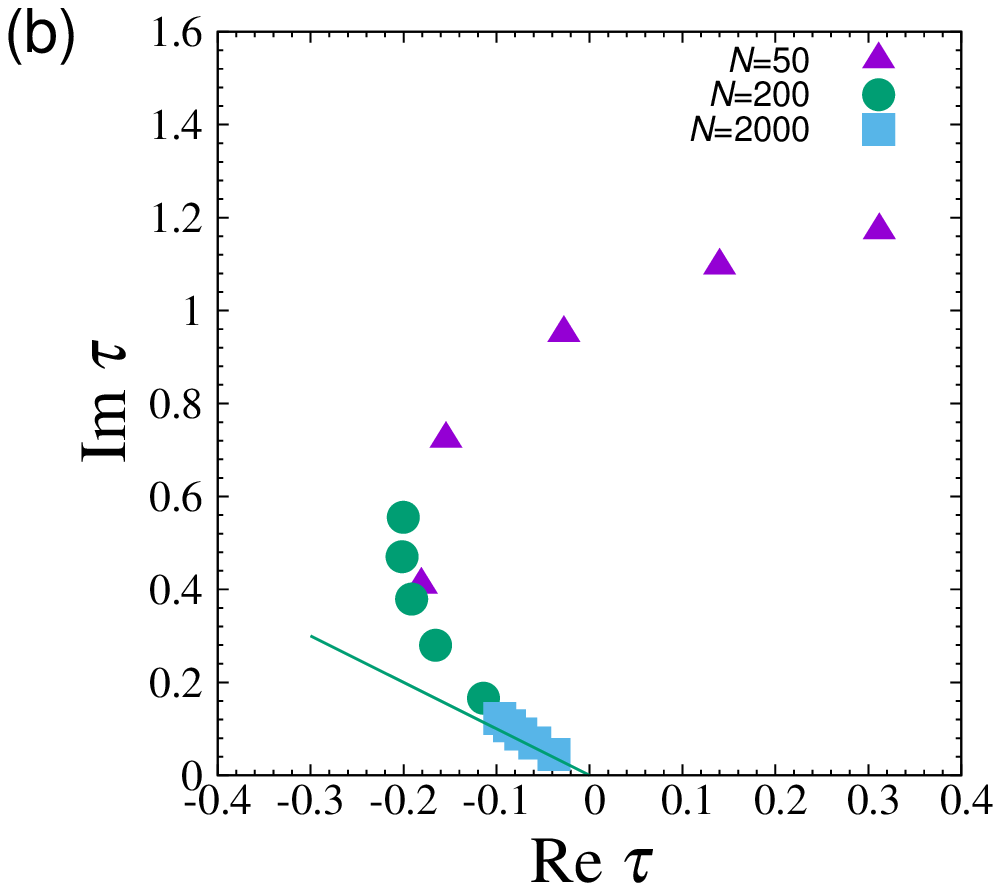}} \caption{First five Fisher zeros
for the approximated partition function for different values of $N$
({a}) in the complex $t$   plane (Eq. (\ref{II.11})) and ({b}) in
the complex $\tau$ plane (Eq. (\ref{II.14})). The solid line
has an impact  angle $\varphi=\pi/4$.\label{fig4}}
\end{figure}

We now turn our attention to the zeros of the approximated partition
function (\ref{II.7}) to check how are these tendencies manifested
in this case. For zero magnetic field, keeping only the term leading
in $1/N$, the partition function (\ref{II.7}) reads
\begin{equation}\label{II.11}
Z_N^{\rm exp}(t)= \int_{-\infty}^{+\infty}\exp\Big(-\frac{x^2
t}{2}-\frac{x^4}{12N}\Big)dx\, .
\end{equation}
The first five Fisher zeros for the approximated partition function
(\ref{II.11}) are shown in the complex $t$   plane in
Fig.~\ref{fig4}{a} for different values of $N$.
Similarly as for the exact partition function (cf.
Fig.~\ref{fig2}{a}), with an increase of $N$, the zeros accumulate
in the vicinity of the critical point and tend to pinch the real
axis at the critical point. An obvious difference in the behaviour
of zeros for the exact and expanded partition functions is that in
the last case the angle of zeros accumulation is very robust and
almost does not depend on $N$. The reason becomes apparent when the
function under the integral in (\ref{II.11}) is rewritten as
\begin{equation}\label{II.12}
Z(z)=\int_{-\infty}^{+\infty}\exp\Big(-z\, x^2 -x^4\Big)dx\, ,
\end{equation}
where the $N$ and temperature dependencies are combined in the
single scaling variable $z$
\begin{equation}\label{II.13}
z=\sqrt{3N}t \, .
\end{equation}
Now it is easy to see that the connection between two sets of zeros
$t_j(N_1)$ and $t_j(N_2)$ calculated  for two different system sizes
$N_1$ and $N_2$ is given by a simple rescaling:
$t_j(N_1)=t_j(N_2)\sqrt{N_1/N_2}$. This, in turn, is manifested by
the fact that zeros calculated for different system sizes align
along the same line in Fig.~\ref{fig4}{a}.

The same expression (\ref{II.11}), {{written}} for different values
of $N$ expressed in terms of the $\tau$ variable takes the form
\begin{equation}\label{II.14}
Z_N^{\rm exp}(\tau)= \int_{-\infty}^{+\infty}\exp\Big(-\frac{x^2
(\tau-\tau^2)}{2}-\frac{x^4(1-\tau)^4}{12N}\Big)dx\, .
\end{equation}
Here, the $\tau$ dependence is non-linear and the zeros  are
non-trivial functions of $N$.  Indeed, the first few are depicted in
Fig.~\ref{fig4}{b} where one finds behaviour rather similar to that
demonstrated in Fig.~\ref{fig2} for the exact partition function:
sets of zeros calculated at different $N$ tend to form the common
curved  locus. However, evaluating (\ref{II.14}) near the critical
point for small $\tau$ and keeping the leading order contributions
in the linear  term  in $\tau$, one arrives at the expression for
$Z_N^{\rm exp}(\tau)$ that coincides with the corresponding
expression (\ref{II.7}) for $Z_N^{\rm exp}(t)$ in which $t$ is
simply substituted by $\tau$. In turn, at zero magnetic field and
close to the critical point, functions $Z_N^{\rm exp}(\tau)$ and
$Z_N^{\rm exp}(t)$ attain the same form given by Eq. (\ref{II.11}).
In particular, in this approximation the $Z_N^{\rm exp}(\tau)$
function was considered in \cite{Glasser86}.

\begin{figure}[t]
\centerline{\includegraphics[angle=0, width=8.5cm]{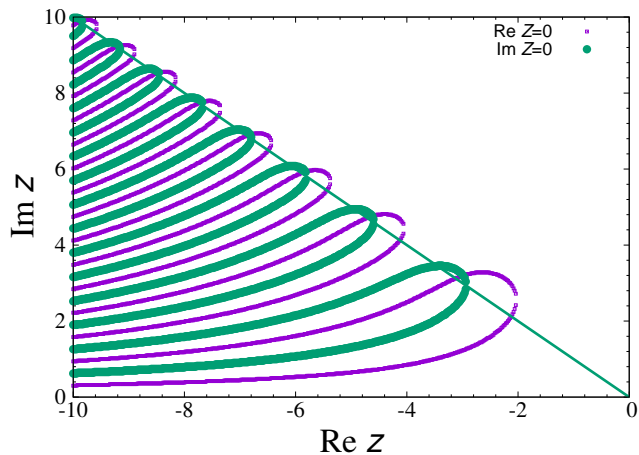} }
 \caption{Loci of zeros for the real and imaginary parts of the
approximated partition function (\ref{II.12}) are plotted as thick
and thin curves respectively (green and violet dots online). The
points where the lines cross give the coordinates of the Fisher
zeros. These have to align along the straight  solid line
in the asymptotic limit. \label{fig5}}
\end{figure}

Zeros of the partition function (\ref{II.12}) in the complex $z$
plane are shown in Fig.~\ref{fig5}. The thick and thin curves (green
and violet online)  give solutions of the equations ${\rm Re}\,
Z(z)=0$ and ${\rm Im}\, Z(z) =0$, respectively. The crossing points
of the lines give coordinates of the Fisher zeros. These have to
align in the asymptotic limit $N\to \infty$ along the line forming
an angle $\varphi=\pi/4$ and starting at the critical point (the
straight  line in the figure).
 As it is clearly seen within the scale of the figure, the higher
the index of the zero, the closer it is to the line.
\begin{figure}[t]
\centerline{\includegraphics[angle=0, width=8cm]{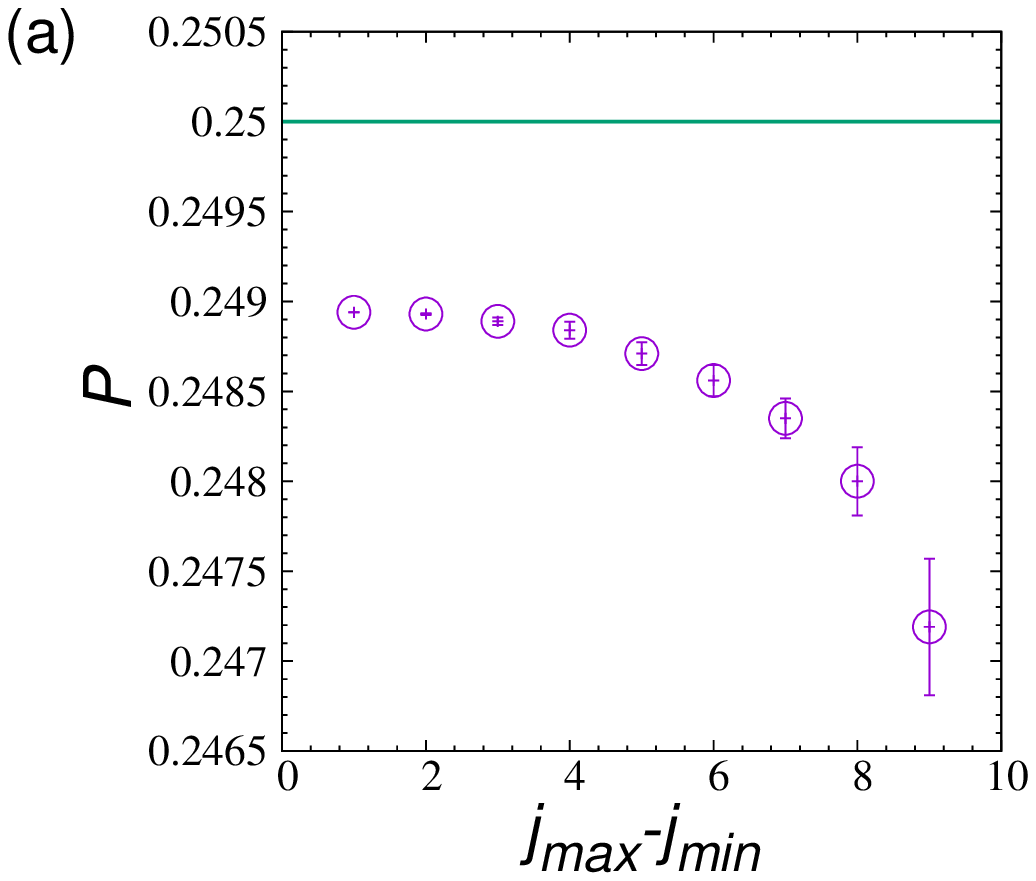}
\includegraphics[angle=0, width=8cm]{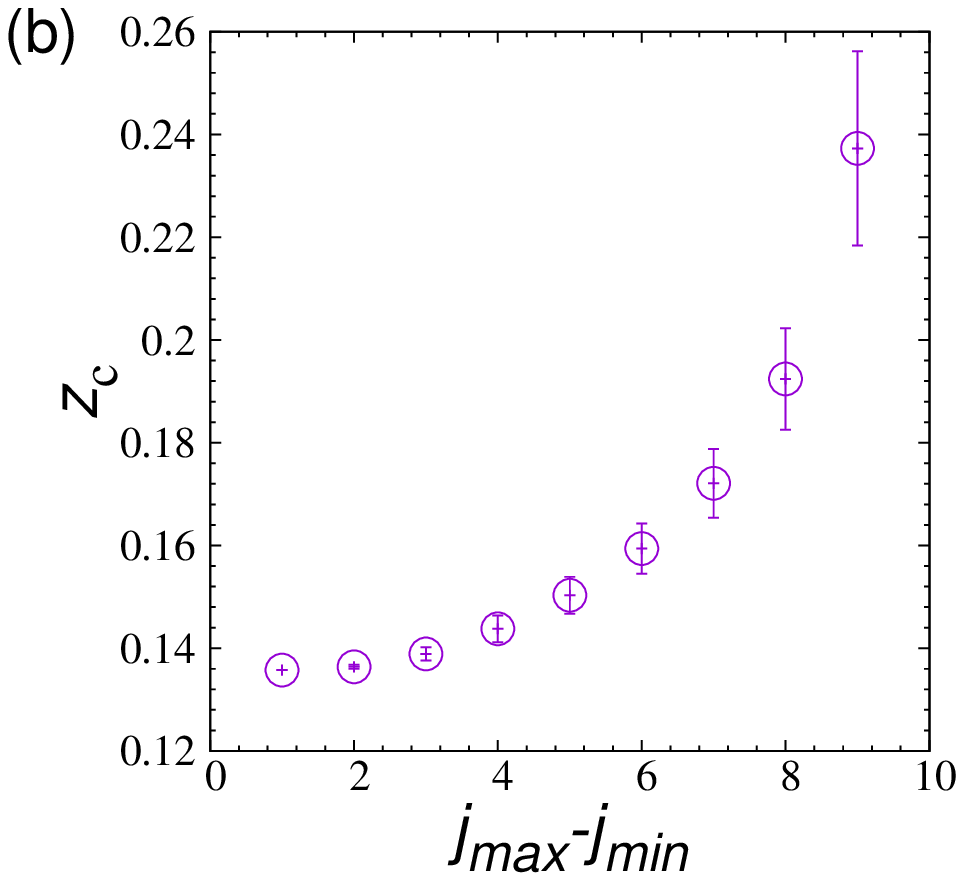}} \caption{Estimates (a) for the
reduced impact angle $P= \varphi/  \pi$  and ({b}) for the critical
temperature $z_c$  for the partition function (\ref{II.12}) obtained
by fitting of Fisher zeros in the interval $j=j_{\rm min},\dots,
j_{\rm max}$ for $j_{\rm max}=10$ and different values of $j_{\rm
min}$. The solid line
 in Panel~a represents the exact value $P=1/4$. \label{fig6}}
\end{figure}
This tendency is further outlined in Fig.~\ref{fig6}, which
demonstrates the behaviour of the angle $\varphi$ and of the
estimates for critical value $z_c$ if they are obtained by fitting
finite numbers of Fisher zeros in the interval $j=j_{\rm min},\dots,
j_{\rm max}$ to a straight line.  Fits for high $j$-values give
better results.  This tendency differs from the one observed above
for the zeros in complex $t$ and  $\tau$ planes (see
{{Fig.~\ref{fig3}}}) and is explained by the form of the scaling
variable (\ref{II.13}), which incorporates now both the temperature
and the system size dependencies.

The coordinates of the first nine Fisher zeros in the $z$   plane as
obtained by numerical solution of the system of equations
(\ref{I.2}) are listed in the left column of  Table~\ref{tab1}.
\begin{table}[b]
\caption{Fisher zeros for the partition function (\ref{II.12}). The
second column shows our numerical results for the zeros given by
$j=1,\dots,9$. In the third column they are compared with the values
given by the asymptotic formula (\ref{II.16}): ${\rm Re}\, z_j
=-{\rm Im}\, z_j =\sqrt{4\pi \, j}$. When digits in the left and
right columns coincide, they are underlined. Starting from $j=9$ the
numerically calculated values coincide with their asymptotic
counterparts within the accuracy presented. \label{tab1}}
\begin{center}
\begin{tabular}{|c|c|c|}
 \hline
  $j$   & Numeric  & Eq. (\ref{II.16})  \\
 \hline
1 & $-\underline{2.98}52+i\underline{3.20}61$   & $-\underline{2.98}23+i \underline{3.20}23$     \\
2 & $-\underline{4.623}6+i\underline{4.770}7$   & $-\underline{4.623}1+i\underline{4.770}0$     \\
3 & $-\underline{5.823}7+i\underline{ 5.941}4$   & $-\underline{5.823}5+i\underline{5.941}1$     \\
4 & $-\underline{6.8167}+i\underline{6.917}6$   & $-\underline{6.8167}+i\underline{6.917}4$     \\
5 & $-\underline{7.682}9+i\underline{7.772}5$   & $-\underline{7.682}8+i\underline{7.772}4$     \\
6 & $-\underline{8.46}10+i\underline{8.542}5$   & $-\underline{8.46}09+i\underline{8.542}4$     \\
7 & $-\underline{9.173}4+i\underline{ 9.2486}$   & $-\underline{9.173}3+i\underline{ 9.2486}$     \\
8 & $-\underline{9.8343}+i\underline{9.904}6$   & $-\underline{9.8343}+i\underline{9.904}5$     \\
9 & $-\underline{10.4536}+i\underline{10.5197}$ & $-\underline{10.4536}+i\underline{10.5197}$   \\
 \hline
    \end{tabular}
\end{center}
\end{table}
As one observes from the table, as the zero-index $j$ increases, its
real and imaginary part become closer, which corresponds to an
approach of $\varphi$ to the value $\varphi=\pi/4$ (see also
Fig.~\ref{fig5}). The value of $z_j$ for high $j$ can be evaluated
asymptotically, writing the partition function (\ref{II.12}) via the
special functions:
\begin{equation}\label{II.15}
Z(z)=\frac{\sqrt{z}}{4}\exp\Big(\frac{z^2}{8}\Big)K_{1/4}\Big(\frac{z^2}{8}\Big)=
\frac{\sqrt{\pi}}{2^{5/4}}\exp\Big(\frac{z^2}{8}\Big)D_{-1/2}\Big(\frac{z}{2^{1/2}}\Big)\,
,
\end{equation}
where $K_{1/4}(x)$ and $D_{-1/2}(x)$ are the Bessel and parabolic
cylinder functions \cite{Rushyk_Grad}, respectively. The leading
term of the asymptotic expansion reads \cite{Glasser86}:
\begin{equation}\label{II.16}
z_j\simeq 2(2\pi j)^{1/2}\exp(3\pi i/4)\Big( 1 + O(1/j)\Big),
  \end{equation}
which in turn leads to the fact that ${\rm Re}\, z_j =-{\rm Im}\,
z_j =\sqrt{4\pi \, j}$ for high $j$. This asymptotic value is shown
for different $j$ in the right column of Table~\ref{tab1}. Equation
(\ref{II.16}) has been obtained in Ref. \cite{Glasser86} for $|z_j|
\gg 1$. Since already for the first zero $|z_1|\simeq 4$, the
asymptotic formula gives quite accurate values for the zeros for all
$j$ as one may see comparing the left and right columns of the
table. Starting from $j=9$ the numerically calculated values
coincide with their asymptotic counterparts within the accuracy
presented.

\subsection{Lee-Yang zeros for the Ising model on the complete graph {{at $T=T_c$}}}\label{IIb}

We now turn to the analysis of the partition function for complex
magnetic field $H$ at $T=T_c$. The exact expression for the
partition function (\ref{II.6}) at $t=\tau=0$ reads:
\begin{equation}\label{II.17}
Z_N (H)= \int_{-\infty}^{+\infty} \exp\Big(-\frac{x^2}{2}+N\ln
\cosh[x/\sqrt{N}+H]\Big)dx.
  \end{equation}
On the other hand, keeping only  contributions leading in $N^{-1}$
in the expanded partition function (\ref{II.7}) one gets:
\begin{equation}\label{II.18}
Z^{\rm exp}_N (H)= \int_{-\infty}^{+\infty} \exp\Big
(-\frac{x^4}{12N} + x\sqrt{N}H)\Big)dx \, .
  \end{equation}

It has been suggested \cite{Itzykson83} that at the critical point
the partition function zeros scale as a fraction of their total
number $j/N$ for large values of the index $j$ . Moreover, many
models give scaling in the ratio $(j - C)/N$ in which $C= 1/2$ is an
empirical fitting factor\cite{Janke01,Banos13}. Recently, a more
comprehensive form for the scaling of the Lee-Yang zeros in the
critical region was suggested \cite{Gordillo13}. In our case it
reads
\begin{equation}\label{II.19}
{\rm Im}\, H_j \, \sim \, \Big( \frac{j-C}{N} \Big)^{\sigma} \, ,
\end{equation}
where the exponent $\sigma$ is related to the order parameter and
heat capacity critical exponents via
\begin{equation}\label{II.20}
\sigma=\frac{\beta\,\delta}{2-\alpha}\, .
\end{equation}
Substituting the values of the exponents (\ref{II.5}) into (\ref{II.20}) one
arrives at
\begin{equation}\label{II.21}
\sigma=3/4\, .
\end{equation}
Next we check whether the scaling form (\ref{II.19}) holds for the Lee-Yang zeros of
the exact and approximated partition functions and we seek to estimate the value of the parameter $C$.

\begin{figure}[t]
\centerline{\includegraphics[angle=0, width=8cm]{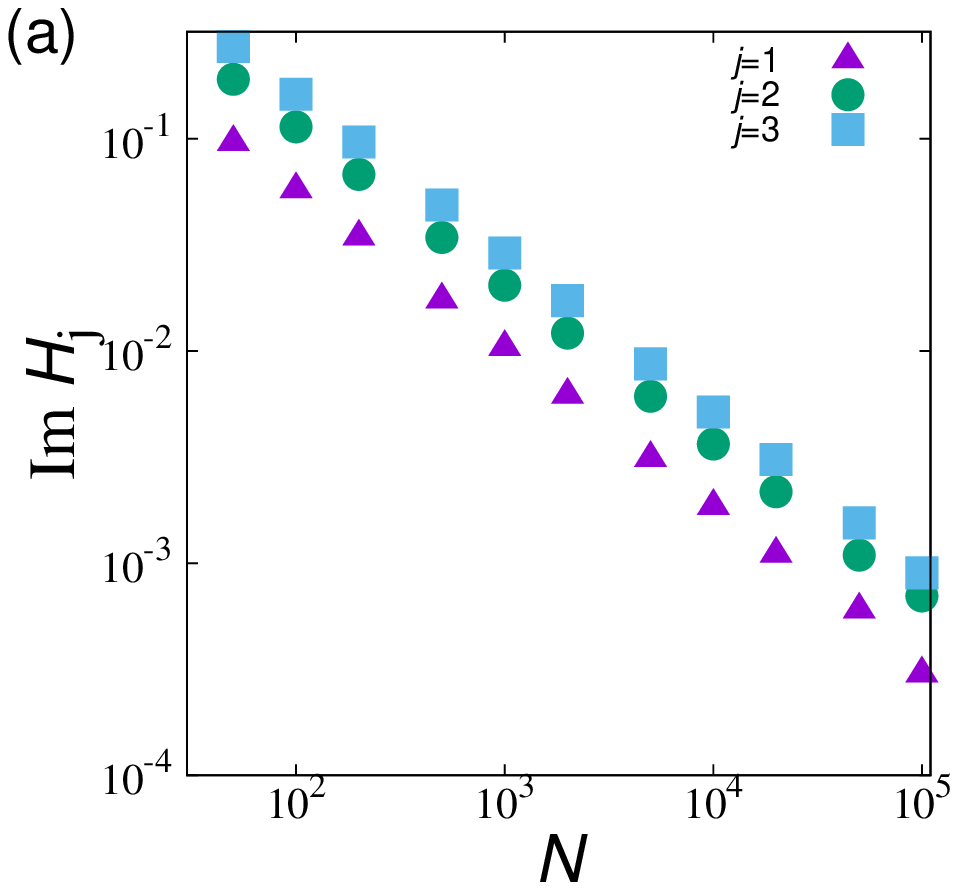}
\includegraphics[angle=0,width=8cm]{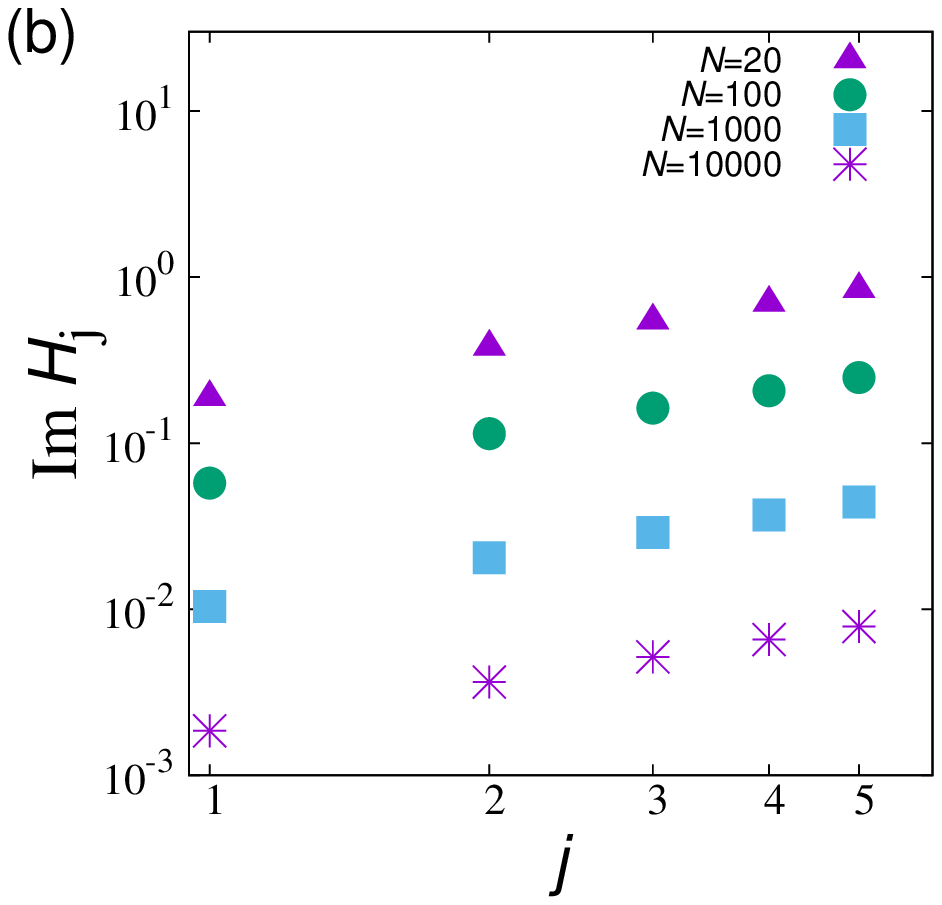}}  \vspace*{1pt}
\centerline{\includegraphics[angle=0,width=8cm]{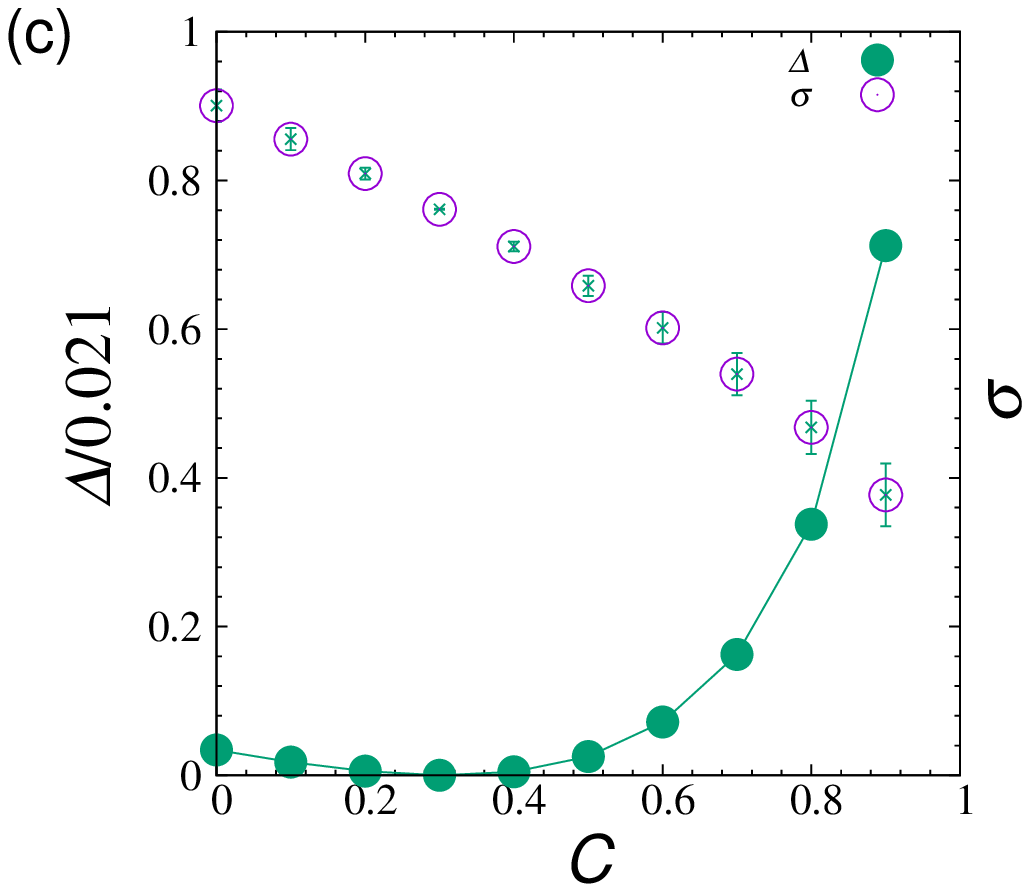}}
 \vspace*{1pt}\caption{The behaviour of ${\rm Im}\,H_j$ for several leading
Lee-Yang zeros of the exact partition function (\ref{II.17}). ({a})
The finite-size ${\rm Im}\, \,H_j(N)$ dependency for first three
zeros $j=1,2,3$. The scaling with $N$, Eq. (\ref{II.19}), holds even
for small $N$ and $j$ with an exponent very close to $\sigma=3/4$.
({b}) Coordinates of the first five zeros calculated for several
values of $N=20,$ 100, 1000, 10000. The scaling exponents remain
almost unchanged for different $N$ but they are far from their
asymptotic value. ({c}) Variance of residuals $\Delta$ and exponent
$\sigma$ obtained while fitting coordinates of the first five Lee
Yang zeros via Eq. (\ref{II.19}) as functions of the fitting
parameter $C$. The optimal $\sigma$ is evaluated at $C_{\rm
opt}=0.31$ where the $\Delta(C)$ curve has minimum. The resulting
value $\sigma(C_{\rm opt})=0.7563(1)$ is close to the exact result
$\sigma=0.75$.  \label{fig7}}
\end{figure}

Solving the system of equations ${\rm Re}\, Z(t,H) =0$ and ${\rm
Im}\, Z(t,H)=0$ at $t=t_c=0$ for the real and imaginary parts of the
exact partition function (\ref{II.17}) we get coordinates of its
zeros in the complex $H$   plane. We find that the Lee-Yang zeros
are purely imaginary in this case. The fact that they lie on the
imaginary axis validates the famous Lee-Yang circle theorem
\cite{LeeYang52}: expressed in terms of the variable $e^H$ all zeros
lie on a circle of unit radius. In Fig.~\ref{fig7}{a} we plot
numerical values of the coordinates of the first three Lee-Yang
zeros as function of the system size $N$. There are two remarkable
feature of the plots of \ref{fig7}{a}: (i) all $H_j(N)$ dependencies
are power laws (represented by straight lines in the log-log plots);
(ii) these power laws are governed by the same value of the exponent
(the lines are parallel) for different $j$. Therefore, the scaling
with $N$ as it is predicted by the equation (\ref{II.19}) holds even
for small $N$ and $j$. Using linear fits for all eleven data points
of the Fig.~\ref{fig7}{a} we find for each $j$: $\sigma=0.749(5)$
($j=1$), $\sigma=0.744(3)$ ($j=2$), $\sigma=0.750(1)$ ($j=3$).

To check how  the scaling of the zeros holds with $j$, we plot in
Fig.~\ref{fig7}{b} coordinates of the first five zeros calculated
for several values of  $N=20,$ 100, 1000, 10000. The fits to the
power law dependency gives the following values of the exponent
$\sigma=0.930(22)$ ($N=20$), $\sigma=0.909(22)$ ($N=100$),
$\sigma=0.901(22)$ ($N=1000$), $\sigma=0.900(21)$ ($N=10000$): the
exponents remain almost unchanged for different $N$ but they are far
from their asymptotic value $\sigma=3/4$, Eq. (\ref{II.21}).
However, introducing a fit parameter $C$ via equation (\ref{II.19})
changes the picture. { We fit the dependence of the first five
Lee-Yang zeros on $j$ on a log-log scale by a linear function at
different values of the fitting parameter $C$ and evaluate goodness
of fits by calculating variance of residuals (reduced $\chi^2$),
i.e. the weighted sum-of-squares residuals divided by the number of
degrees of freedom, further denoted by $\Delta$. The $\Delta(C)$
curve is shown in Fig.~\ref{fig7}{c}, the optimal value of the
fitting parameter $C_{\rm opt}$ corresponds to minimum of $\Delta$.
In our case $C_{\rm opt}=0.31$. Therefore, the value of the exponent
$\sigma$ is calculated as $\sigma(C_{\rm opt})=0.7563(1)$ and is
close to the exact result $\sigma=0.75$.}

\begin{figure}[t]
\centerline{\includegraphics[angle=0, width=8.5cm]{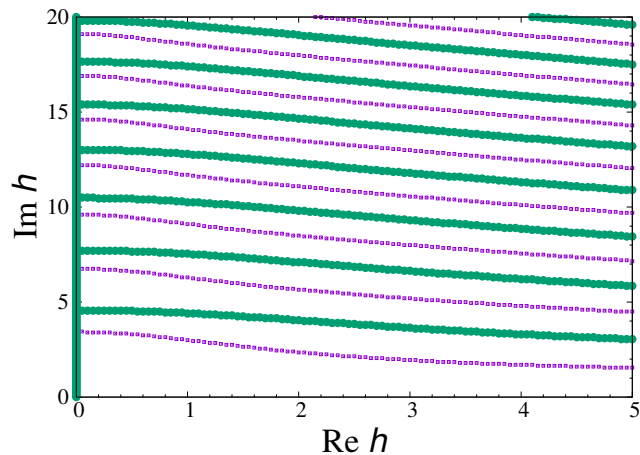}}
\caption{Lines of zeros for the real and imaginary part of the
approximated partition function (\ref{II.22}) at $T=T_c$ in the
complex magnetic field plane, thin  and thick curves (violet and
green online) respectively. The points where the lines of different
colour cross give the coordinates of the Lee-Yang zeros. Note that
one of the ${\rm Im}\, Z(h) = 0$ lines coincides with the vertical
axis in the plot. \label{fig8}}
\end{figure}

For the expanded partition function, Eq. (\ref{II.18}) can be conveniently written
in terms of the rescaled variables:
\begin{equation}\label{II.22}
Z(h)= \int_{-\infty}^{+\infty} \exp \Big( -x^4 -x\, h \Big ) dx,
\end{equation}
where the field and the system size dependence is absorbed into a single variable
\begin{equation}\label{II.23}
h=H(12N^3)^{1/4}.
\end{equation}

Fig.~\ref{fig8} displays  the lines of zeros for the real and
imaginary parts of the partition function (\ref{II.22}) at $T=T_c$
in the complex magnetic field $h$   plane, thin and thick curves
(violet and green online) respectively.
 Again, the coordinates of the Lee-Yang
zeros (the crossing points of the lines) are purely imaginary
(obeying the Lee-Yang circle theorem). The scaling of $h_j$ with the
system size $N$ is given by (\ref{II.23}) and hence the scaling
exponent is equal to its asymptotic value (\ref{II.21}) for any
index  $j$. Note, that since the real part of the Lee-Yang zero
coordinate ${\rm Re}\, h =0$, it is straightforward to show from
(\ref{II.18}) that the  imaginary coordinate ${\rm Im}\,{h}$ is
obtained as a solution of an integral equation:
\begin{equation}\label{II.24}
\int_0^\infty e^{-x^4}\cos(x {\rm Im}\,{h}) \, dx=0\, .
\end{equation}

\begin{figure}[t]
\centerline{\includegraphics[angle=0, width=8cm]{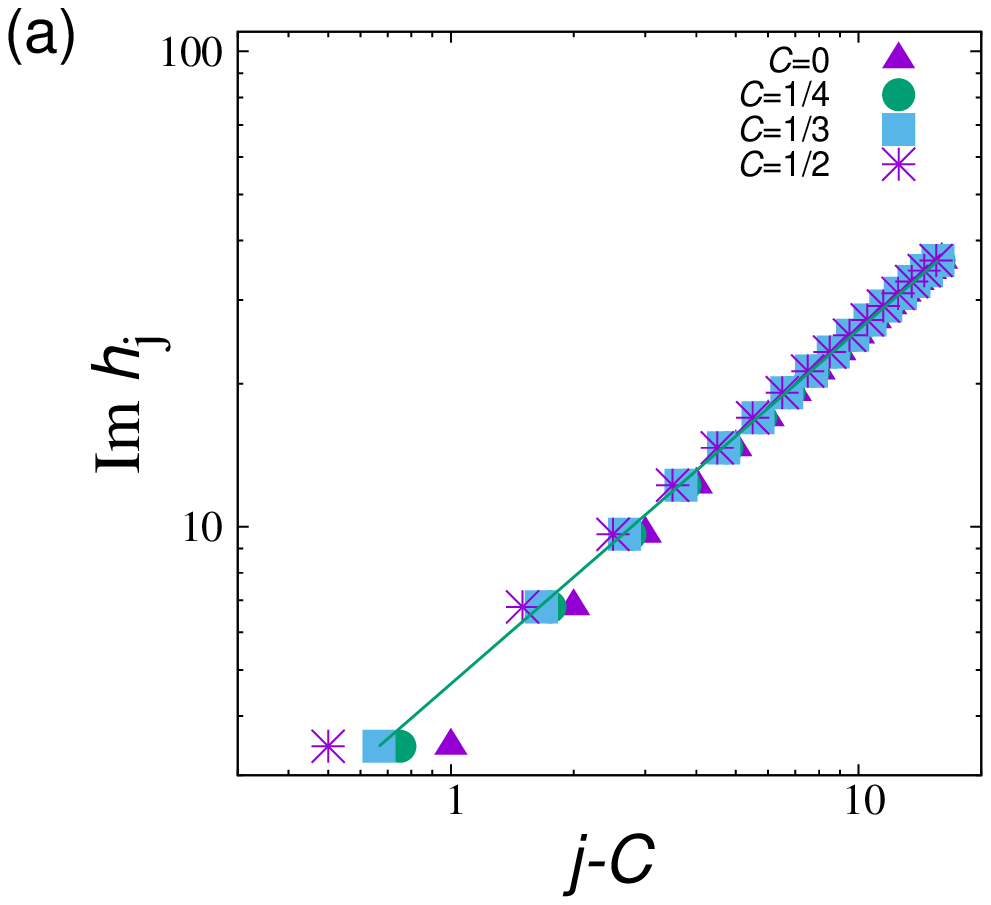}
\hspace{1em}
\includegraphics[angle=0, width=8cm]{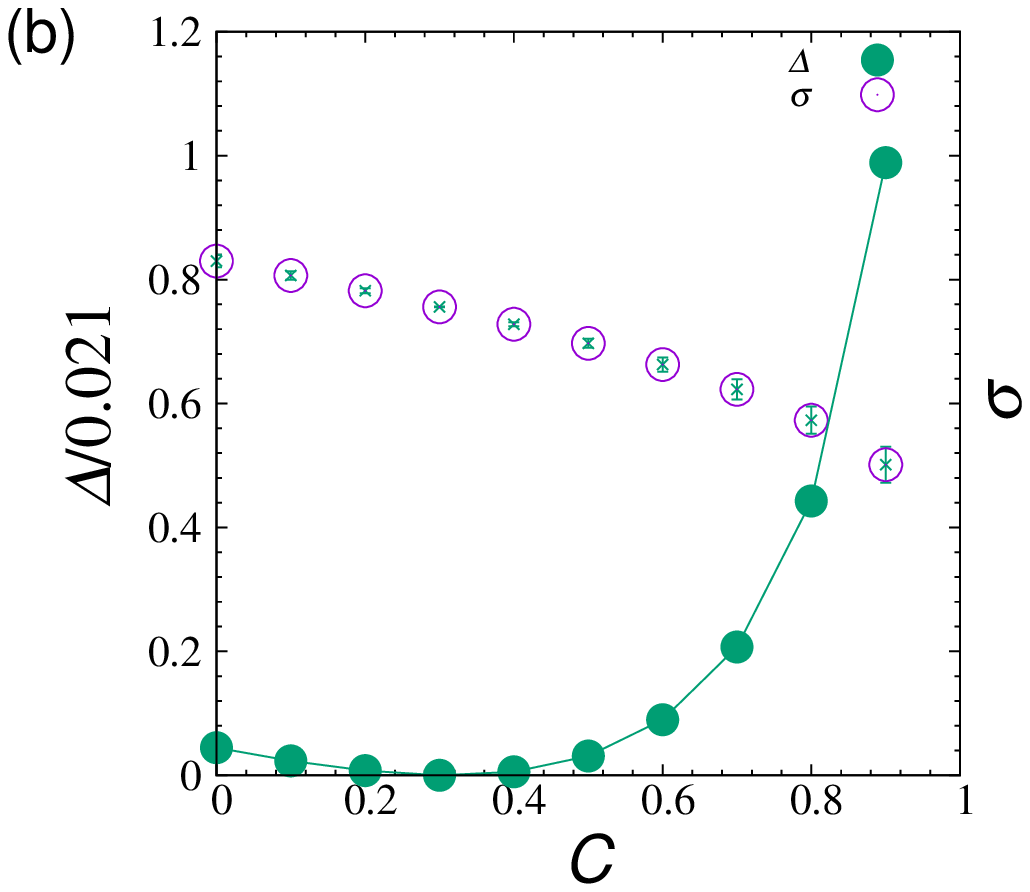}}
\caption{The behaviour of ${\rm Im}\,h_j$ for leading Lee-Yang zeros
of the approximated partition function (\ref{II.22}). ({a})
Coordinates ${\rm Im}\, h_j $ of numerically calculated first
sixteen  Lee-Yang zeros as functions of  $j-C$ for different values
of $C=0, \, 1/4,
\, 1/3, \, 1/2$. The solid line 
 is an eye guide to
show linear scaling with an exponent $\sigma=3/4$. Such scaling is
manifested by a merge of the data points with an increase of $j$.
({b}) Variance of residuals $\Delta$ and exponent $\sigma$ obtained
while fitting coordinates of the first sixteen Lee Yang zeros via
Eq. (\ref{II.19}) as functions of the fitting parameter $C$. The
value of the exponent $\sigma$ calculated at $\sigma(C_{\rm
opt})=0.7531(2)$ is close to the exact result $\sigma=0.75$.
\label{fig9}}
\end{figure}

Let us consider now  the scaling with $j$. To this end, in
Fig.~\ref{fig9} we plot numerically the coordinates ${\rm Im}\, h_j
$ of the first sixteen  Lee-Yang zeros  as functions of  $j-C$ for
different values of $C=0, \, 1/4, \, 1/3, \, 1/2$. As expected, for
large enough $j$ the linear scaling with $j$ occurs and the data
points align along a single line with the tangent $\sigma\simeq3/4$.
However, as for the exact partition function, this dependence is
non-linear for smaller values of $j$ and the choice of an
appropriate fitting parameter $C$ { similar as it was done for the
exact function, see Fig.~\ref{fig7}{c},}  allows to improve the
linearity. { As a result, we get the optimal value $C_{\rm
opt}=0.31$ and $\sigma(C_{\rm opt})=0.7531(2)$.}

\subsection{Motion of the Fisher zeros in a real external field}\label{IIc}

In the final part of this section we  analyze the motion of the
Fisher zeros in the presence of a (real) external field \cite{Itzykson83}. Since the Lee-Yang zeros
of the model under consideration have purely imaginary coordinates,
in the vicinity of the critical point the lines of motion of Fisher
zeros form an angle $\psi$ that encodes order parameter critical
exponents via Eq.~(\ref{I.12}). In our case,
for the values of the exponents $\beta$ and $\delta$ given by
(\ref{II.5}) one gets for the angle:
\begin{equation}\label{II.26}
\psi=\frac{\pi}{3}\, .
\end{equation}
The expression for the approximated partition function in the vicinity of the critical point
is obtained from (\ref{II.7}) and reads
\begin{equation}\label{II.27}
Z^{\rm exp}_N (t,H)= \int_{-\infty}^{+\infty} \exp\Big (\frac{-t\,
x^2}{2}-\frac{x^4}{12N} + x\sqrt{N}H \Big)dx.
\end{equation}
To get (\ref{II.27})  we have kept, as before, only the leading
order contributions in $H$ and $t$ terms. As  was
pointed above, in this case the approximated expressions for the
partition function attain the same form both in $t$- and $\tau$-
representations. Therefore,  the analysis of the Fisher zeros
presented below concerns both the complex $t$- and complex
$\tau$-planes equally well. The expansion (\ref{II.27}) can be
conveniently rewritten in the rescaled variables $z$, $h$, Eqs.
(\ref{II.13}), (\ref{II.23}):
\begin{equation}\label{II.27A}
Z^{\rm exp} (z,h)=\int_0^\infty e^{-zx^2-x^4+hx}dx.
\end{equation}
Note that for $h=0$ or $z=0$ (\ref{II.27A}) reduces to Eq.
(\ref{II.12}) or (\ref{II.22}) respectively.

\begin{figure}[t]
\centerline{
\includegraphics[angle=0,width=8cm]{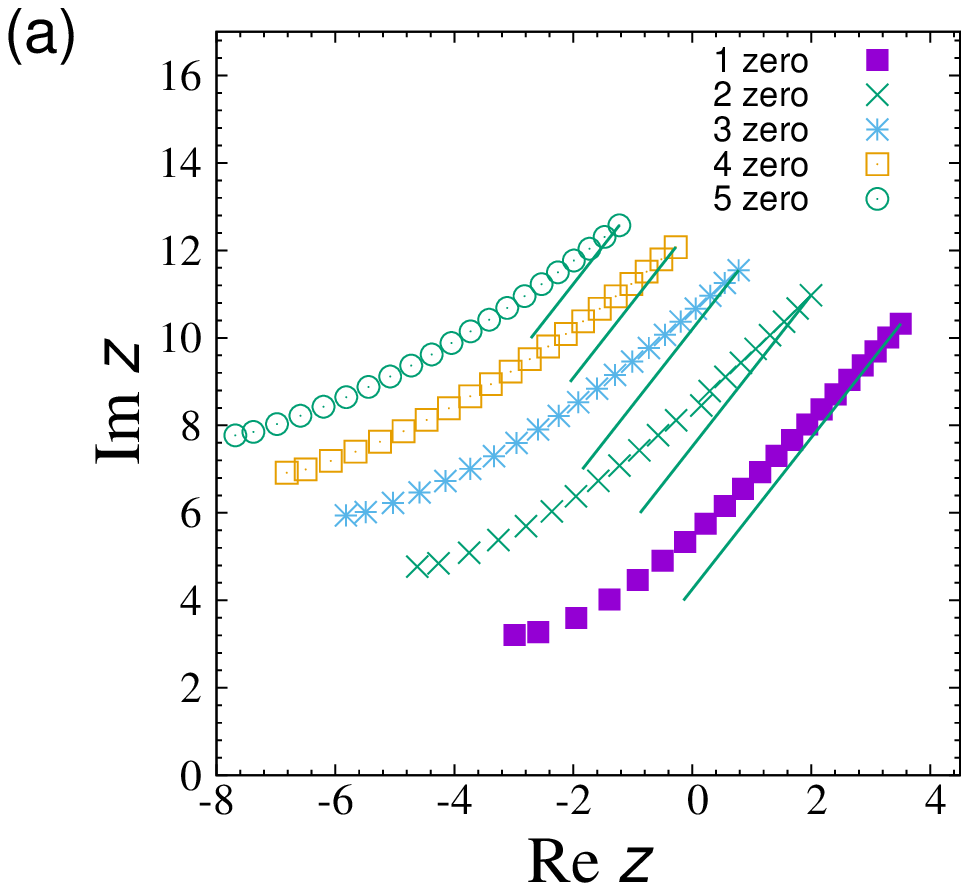}
\includegraphics[angle=0,
width=8cm]{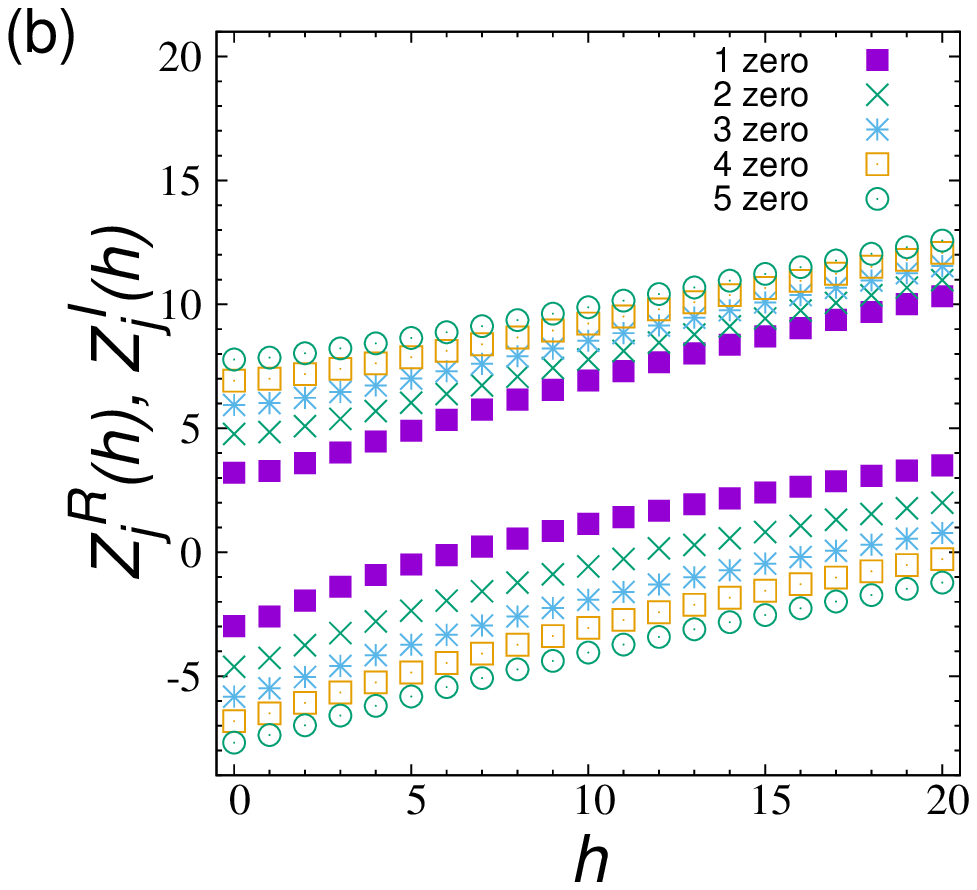}}
 \vspace*{1pt}
\caption{({a}) Coordinates of the first five Fisher zeros of the
partition function (\ref{II.27A}) in the complex $z$   plane for
different values of the magnetic field $h_j=j$, $j=0,1,\dots,20$.
The plots have a tendency to settle along lines 
forming an angle $\psi=\pi/3$ with a real $z$   axis.
 ({b}) Scaling functions for the first five zeros ${\cal Z_{\it j}^{\,R}}\, (h)$ (lower plots),
 ${\cal Z_{\it j}^{\,I}}\, (h)$
 (upper plots) for the real and imaginary part of the first Fisher zero coordinate,
  Eq. (\ref{II.34A}),
as functions of the scaling variable
$h=HN^{1+\frac{\beta}{\alpha-2}}$. \label{fig10}}
\end{figure}

Fig.~\ref{fig10}{a} shows the coordinates of the first five Fisher
zero for different values of the (real) magnetic field in the
complex $z$ plane. The coordinates have been calculated for
different values of the magnetic field $h_j=j$, $j=0,1,\dots,20$.
For $h_j=0$ we recover the values of the coordinates shown in
Fig.~\ref{fig5}.  One can see the tendency of the zeros to settle
along a line forming an angle $\psi=\frac{\pi}{3}$ with the real $z$
axis (shown by a solid line in the figure). However, the asymptotics
are preceded by a crossover region for very small values of $h$.
Similar tendency for a first few zeros of a 3D Ising model was
observed in Ref. \cite{Itzykson83} and explained by the finite-size
corrections \cite{Zuber83}. The closer the zeros  to the critical
point, the smaller the crossover region.

To further analyze motion of the zeros we apply the finite size
scaling (FSS) as outlined below. According to the FSS theory
\cite{FSS}, for a $d$-dimensional system in the vicinity of the
critical point one expects the following scaling for the real and
imaginary parts of the $j$th zero coordinate in the reduced
temperature $t$   plane:
\begin{eqnarray}\label{II.28}
{\rm Re}\,\,  t_j (N, H) &=& b^{-1/{\nu}}{\cal T_{\it j}^{\,R}}\,
(H\,b^{y_H}, N\, b^{-d})\, , \\
\label{II.29} {\rm Im}\,\, t_j (N, H)&=& b^{-1/{\nu}}{\cal T_{\it
j}^{\, I}}\, (H\, b^{y_H}, Nb^{-d})\, .
\end{eqnarray}
Here, $b$ is the scaling factor, $\nu$ and $y_H$ are the correlation
length critical exponent and field scaling dimension, ${\cal T_{\it
j}^{\,R}}\, (x,\,y))$ and ${\cal T_{\it j}^{\,I}}\, (x,\,y)$ are the
scaling functions. Being generalised homogeneous functions of two
variables, the scaling functions can be rewritten as functions of a
single conveniently chosen scaling variable. Choosing the factor
$b=N^{-1/d}$ the expressions (\ref{II.28}), (\ref{II.29}) attain the
following form:
\begin{eqnarray}\label{II.30}
{\rm Re}\,\,  t_j (N,H) &=& N^{-1/{d\nu}}{\cal T_{\it j}^{\,R}}\, (H\,N^{-y_H/d},\, 1)\, , \\
\label{II.31} {\rm Im}\,\, t_j (N,H)&=& N^{-1/{d\nu}}{\cal T_{\it
j}^{\, I}}\, (H\,N^{-y_H/d},\, 1)\, .
\end{eqnarray}
making use of the familiar (hyper)scaling relations ($d\nu=2-\alpha$,
$y_h/d=1-\beta/{2-\alpha}$) we get the following expressions for the motion of zeros:
\begin{eqnarray}\label{II.32}
{\rm Re}\,\,  t_j (N,H) &=& N^{\frac{1}{\alpha-2}}\, {\cal T_{\it j}^{\,R}}\, (H\,N^{1+\frac{\beta}{\alpha-2}})\, , \\
\label{II.33} {\rm Im}\,\, t_j (N,H)&=& N^{\frac{1}{\alpha-2}}\,
{\cal T_{\it j}^{\,I}}\, (H\,N^{1+\frac{\beta}{\alpha-2}})\,  ,
\end{eqnarray}
where we have introduced the single variable functions:
\begin{equation}\label{II.34}
{\cal T_{\it j}^{\,R}}\, (x)\equiv {\cal T_{\it j}^{\,R}}\, (x,\, 1),\hspace{2em}
{\cal T_{\it j}^{\,I}}\, (x)\equiv {\cal T_{\it j}^{\,I}}\, (x,\, 1)\, .
\end{equation}

Note that relations (\ref{II.32})-(\ref{II.33}) contain the system
size only via the number of particles $N$ (and not via the spacial
extent and dimension) and therefore are convenient for using them
for systems on graphs, where the notion of Euclidean dimension is
not defined. Equations (\ref{II.32}), (\ref{II.33}) can be rewritten
in the rescaled variables $z$, $h$ (see Eqs.(\ref{II.13}) and
(\ref{II.23})):
\begin{eqnarray}\nonumber
{{\rm Re}\, z_j}&=&{\cal Z_{\it j}^{\,R}}\, (h)\, ,\\\label{II.34A}
 {{\rm Im}\,
z_j}&=& {\cal Z_{\it j}^{\,I}}\, (h)\, ,
\end{eqnarray}
where
\begin{equation}\label{II.34B}
{\cal Z_{\it j}^{\,R}}\, (h) \equiv \sqrt{3}\, {\cal T_{\it
j}^{\,R}}\, (h/\sqrt[4]{12})\, ,\hspace{2em} {\cal Z_{\it
j}^{\,I}}\, (h)\equiv \sqrt{3}\, {\cal T_{\it j}^{\,I}}\,
(h/\sqrt[4]{12})\, .
\end{equation}
We plot the scaling functions (\ref{II.34B}) in Fig.~\ref{fig10}{b}
for the first five zeros $j=1,\dots,5$. The prominent feature of the
plot is that a ratio of the values of the scaling functions at $h=0$
gives the value of the Fisher pinching angle:
\begin{equation}\label{II.35}
{\cal T_{\it j}^{\,I}}\, (0)/{\cal T_{\it j}^{\,R}}\, (0) = {\cal
Z_{\it j}^{\,I}}\, (0)/{\cal Z_{\it j}^{\,R}}\, (0)=\tan \varphi\, .
\end{equation}
Indeed, with the value $\tan\varphi=\tan\pi/4=-1$, Eq.
(\ref{II.10}), one gets ${\cal T_{\it j}^{\,I}}\, (0)=-{\cal T_{\it
j}^{\,R}}\, (0)$ or ${\cal Z_{\it j}^{\,I}}\, (0)=-{\cal Z_{\it
j}^{\,R}}\, (0)$ as is nicely observed in the figure.

\section{Ising model on an annealed scale-free network}\label{III}

Now, with  knowledge of the behaviour of the partition function
zeros on a complete graph to hand, we consider the critical
behaviour of the Ising model on complex networks, i.e. on a random
graph \cite{networks}. The Hamiltonian of the model in this case
reads
\begin{equation}\label{III.1}
-{\cal H}=\frac{1}{2}\sum_{l\neq m}J_{lm}S_lS_m +H \, \sum _l S_l \,
.
\end{equation}
Here, the sums are performed over all graph nodes $l,m$ and the
adjacency matrix $J$ stores all information about the graph
structure: the matrix elements $J_{lm}=1$ if the nodes are linked
and $J_{lm}=0$ otherwise. In random graphs different nodes have
different number of links (different node degrees $K$). The node
degrees are random variables and inherent features of their
distribution $p(K)$ appear to play one of the key roles governing
universal critical behaviour \cite{Dorogovtsev08}. As a first
instance of a complex network we will consider an {\em annealed}
network.
This network is defined as an ensemble of all networks
consisting of $N$ nodes with a given degree sequence
$\{K_1,K_2,\dots,K_N\}$, maximally random under the constraint that
their degree distribution is a given one, $p(K)$ (see e.g.
\cite{Lee09}).
Such a construction resembles the uncorrelated
configurational model of a complex network (see e.g.
\cite{Dorogovtsev02}). However the latter is an example of a quenched
network, whereas for the Ising model on an annealed network, the
graph configuration is also fluctuating just like Ising spins do.
When thermodynamic properties are calculated, the presence of
quenched disorder is taken into account by averaging  the free
energy over different disorder configurations, whereas in the
annealed case the partition function is averaged \cite{Brout59}.
Therefore, considering the Ising model on an annealed network, we
will be interested in the behaviour of the partition function
averaged with respect to different network configurations.

\subsection{Partition function}\label{IIIa}

For the Hamiltonian (\ref{III.1}) the corresponding partition function is obtained by:
\begin{equation}\label{III.2}
Z_N (T,H)={\rm Tr}_{\,S} \, {\rm Tr}_{\, J} \, e^{-\beta \cal{H}}\,
.
\end{equation}
As in (\ref{II.2}), the first trace is taken over the spin system:
${\rm Tr}_{\,S}\, (\dots)=\prod_{l}\sum_{S_l=\pm 1}\, (\dots)$,
whereas the second one means an averaging with respect to the
distribution of the network links ${\cal P}(J)$: ${{{\rm
Tr}}}_{\,J}\, (\dots) =\prod_{l,m, l \neq m}\sum_{J_{lm}=0,1} \,
{\cal P}(J) \, (\dots)$.

To construct an annealed network of $N$ nodes $l=1,\dots,N$, each
node $l$ is assigned a label $k_l$ and the probability of a link
between nodes $l$ and $m$ is defined as:
\begin{equation}\label{III.3}
p_{lm}=\frac{k_lk_m}{N\langle k \rangle}\, ,
\end{equation}
where $\langle k \rangle=\frac{1}{N}\sum_{l}k_l$. The variables $k$
are taken from the distribution $p(k)$ and indicate the expected
node degree. Indeed, it is straightforward to show that the expected
value of the node degree $\mathbb{E} K_l = \sum_m p_{lm} = k_l$.
One
can show \cite{Lee09} that averaging (\ref{III.2}) over the
distribution of network links with probability function
(\ref{III.3}) leads to the following expression for the partition
function:
\begin{equation}\label{III.4}
Z_N(T,H)={{\rm Tr}_{\,S}} \, \exp\left( \frac{1}{2N\langle k \rangle
T}\sum_{l\neq m}S_lS_m k_lk_m +\frac{H}{T} \, \sum _l S_l \right)\,
.
\end{equation}
An analogous expression for the partition function is usually
obtained also within the mean field approximation for the Ising
model on an uncorrelated quenched network \cite{Dorogovtsev02,
Dorogovtsev08}(configurational model). For the annealed network
however the factor in front of the double sum in (\ref{III.4}) is a
certain function of temperature \cite{Lee09}. Since in our study we
are interested in the angles of incidence of partition function
zeros, which are independent of any real analytic parametrization of
the temperature plane, we keep in (\ref{III.4}) only a linear term
of this function.

Since for the annealed network the interaction term in (\ref{III.1}) attains a separable form, one can
apply the Stratonovich-Hubbard transformation to (\ref{III.4}) to take the trace exactly and to get the
following expression for the partition function:
\begin{equation}\label{III.5}
Z_N (T,H)= \int_{-\infty}^{+\infty} \exp\Big (\frac{-N\langle k
\rangle x^2}{2T}+\sum_{l}\ln \cosh[( xk_l+H)/T]\Big)dx \, .
\end{equation}
As in the previous section,  the prefactors are
omitted here and below. Now, the sum over $l$ in the exponent (\ref{III.5}) can be
rewritten in terms of the integral over $k$ for a given distribution
function $p(k)$:
\begin{equation}\label{III.6}
\sum_lf(k_l)=N\int_{k_{\rm min}}^{k_{\rm max}}p(k)f(k)dk\, ,
\end{equation}
where $k_{\rm min}$ and $k_{\rm max}$ are the minimal and maximal values of the variable $k$.
For further analysis it will be convenient to keep the integral in $x$ in the positive half-plane leading to
the  representation for the partition function,
\begin{eqnarray}\nonumber
Z_N (T,H)&=& \int_{0}^{+\infty} e^{\frac{-\langle k \rangle
x^2T}{2}} \Big \{ \exp \Big [ N \int_{k_{\rm min}}^{k_{\rm max}} dk
p(k)\ln \cosh\Big(\frac{x k} {\sqrt{N}}+\frac{H}{T} \Big)\Big ] +
\\ \label{III.7} &&
\exp \Big [ N \int_{k_{\rm min}}^{k_{\rm max}} dk p(k)\ln
\cosh\Big(- \frac{x k} {\sqrt{N}}+\frac{H}{T} \Big)\Big ] \Big
\}\,dx\, .
\end{eqnarray}
For the complete graph of $N$ nodes substituting in (\ref{III.7}) $p(k)=\delta(k-N+1)$ one recovers (\ref{II.6}).
In this section we are interested in the scale-free networks, when the function $p(k)$
is given by a power law
\begin{equation}\label{III.8}
p(k)=c_\lambda\,k^{-\lambda}\, ,
\end{equation}
with a normalizing constant $c_\lambda$. Note, that scale-free
networks with $k_{\rm min}=1$ do not possess a spanning cluster for
$\lambda > \lambda_c$ ($\lambda_c=4$ for continuous degree
distribution and $\lambda_c\simeq 3.48$ for the discrete one
\cite{Aiello00}). To avoid this restriction, without loss of
generality  we choose from now on $k_{\rm min} =2$,  whereas for the
upper integration boundary in (\ref{III.6}) we take in the
thermodynamic limit $\lim_{N\to \infty}k_{\rm max}\to
\infty$.\footnote{The leading in $N$ value of this integral
representation does not depend on the way in which $k_{\rm max}$  tends
to infinity.
These are the next-leading terms that will depend on
the $N$ dependency of the upper cut-off. See Appendix A for more
details.} Then, for the scale-free network, Eq. (\ref{III.7}) can be
conveniently rewritten as:
\begin{equation} \label{III.9}
Z_N (T,H)= \int_{0}^{+\infty} e^{\frac{-\langle k \rangle x^2T}{2}}
\Big \{ \exp \Big [I^+_\lambda(x)\Big ] + \exp \Big
[I^-_\lambda(x)\Big ] \Big \}\,dx\, ,
\end{equation}
where
\begin{equation}\label{III.10}
I^\pm_\lambda(x)= c_\lambda \Big(\frac{x}
{\sqrt{N}}\Big)^{\lambda-1}N\, \int_{\frac{2x}{\sqrt{N}}}^{\infty}
 \frac{1}{y^\lambda}\ln \cosh\Big(\pm y +\frac{H}{T} \Big) \, dy \, .
\end{equation}
Consider the integral in (\ref{III.10}).  For large $y$ the integrand decays as $y^{1-\lambda}$ and the integral is
finite at the upper integration boundary for the values $\lambda>3$
we are interested in.
However, for small $y$ the integrand behaves as $(\pm y+H/T)^2y^{-\lambda}+\dots$ and leads
to divergent terms when the lower integration boundary is set to
zero, i.e. in the thermodynamic limit $N\to\infty$. These divergent
terms do not appear in the whole expression since they are canceled
by the $N$-dependent prefactor in (\ref{III.10}). To single them out
and to show this cancelation explicitly, it is instructive to
consider the function  $I^\pm_\lambda(x)$ (\ref{III.10}) for
different values of $\lambda$ (see e.g. \cite{Krasnytska13} for a
more detailed account). As in the former section, we make asymptotic
estimates for small $H$. Leading terms of the resulting expressions
read:
\begin{eqnarray}\label{III.15}
I^\pm_\lambda(x)&=& N \Big [\frac{\langle k^2 \rangle}{2}
\frac{x^2}{N}-a(\lambda)
\Big(\frac{x}{\sqrt{N}}\Big)^{\lambda-1}\,\pm \, \frac{\langle k
\rangle H x}{T \sqrt{N}} \Big] \, , \hspace{2.1em} 3<\lambda<5\, , \\
\label{III.16} I^\pm_\lambda(x)&=&N \Big [ \frac{\langle k^2
\rangle}{2}\, \frac{x^2}{N}- \frac{\langle k^4
\rangle}{12}\Big(\frac{x}{\sqrt{N}}\Big)^{4} \pm \,  \frac{\langle k
\rangle H x}{T \sqrt{N}} \Big] \,  , \hspace{5.0em} \lambda>5\, ,
\end{eqnarray}
where the numerical values of the coefficients
$a(\lambda)=-c_\lambda\int_0^\infty dy\, y^{-\lambda}(\ln\cosh y
-y^2/2)$, $a(\lambda)>0$,  are listed for different $\lambda$ in
\cite{Krasnytska13} and the moments of the variable $k$ are
calculated with the distribution (\ref{III.8}).
The case $\lambda=5$
is to be considered separately. Integrating out the logarithmic
singularity one gets for the first leading terms:
\begin{equation}\label{III.17}
I^\pm_\lambda(x)= N \Big [\frac{\langle k^2 \rangle}{2}\,
\frac{x^2}{N}- \Big(\frac{x}{\sqrt{N}}\Big)^{4}\frac{\ln N}{24}\pm
\, \frac{\langle k \rangle Hx}{T \sqrt{N}} \Big] \, .
\end{equation}

With the expressions (\ref{III.15})--(\ref{III.17}) for
$I^\pm_\lambda(x)$ we are in a position to analyse the partition
function (\ref{III.9}) at different values of $\lambda$. As is well
established by now, critical behaviour of a system on a scale-free
network with the partition function (\ref{III.9}) depends of the
value of the exponent $\lambda$ in an essential way \cite{networks,
Dorogovtsev08}. In particular, the system remains ordered for any
finite temperature at $\lambda\leq 3$. A second order phase
transition at finite temperature occurs for $\lambda>3$. It is
governed by standard mean field exponents (\ref{II.5}) in the region
$\lambda\geq5$ with logarithmic corrections at $\lambda=5$, however
the exponents attain $\lambda$-dependency for $3<\lambda<5$
\cite{Dorogovtsev02}:
\begin{equation}\label{III.11}
\alpha=(\lambda-5)/(\lambda-3) , \hspace{0.5em} \beta=1/(\lambda-3), \hspace{0.5em} \delta=\lambda-2, \hspace{0.5em}
\gamma=1\, .
\end{equation}
{{ The mean-field approximation delivers the classical}} value for
the magnetic susceptibility exponent $\gamma=1$ for any $\lambda>3$.
The other exponents however become $\lambda$ dependent at
$3<\lambda<5$. This is because the Ising model on complex scale-free
networks follows a heterogeneous rather than homogeneous mean-field.

Our task will be to describe the phase transition in the Ising model
on an annealed scale-free network in terms of the partition function
zeros, similarly as it was done for the complete graph in the
previous section. Substituting values of the exponents
(\ref{III.11}) together with the critical amplitude ratio
$A_+/A_-=0$ \cite{Palchykov10} into relations (\ref{I.3}),
 (\ref{II.20}) we get for the region $3<\lambda<5$:
\begin{equation}\label{III.12}
\varphi=\frac{\pi(\lambda-3)}{2(\lambda-1)},
 \end{equation}
 \begin{equation}\label{III.14}
\sigma=\frac{\lambda-2}{\lambda-1}.
 \end{equation}
 Let us check if these values can be obtained directly from the
analysis of the zeros of the partition function (\ref{III.9}) in
complex $T$ and $H$  planes.

\subsection{Fisher zeros for the Ising model on an annealed scale-free network at $H=0$}\label{IIIb}

Substituting expressions for $I^\pm_\lambda(x)$
(\ref{III.15})--(\ref{III.17}) at $H=0$ into the partition function
(\ref{III.9}) at different values of $\lambda$ we get:
\begin{eqnarray}\label{III.18}
 Z_N(t)=\left\{
\begin{array}{lcc}
                \int_{0}^{+\infty} \exp{\Big[-\frac{\langle k^2 \rangle x^2t}{2}-a(\lambda)N
                \Big(\frac{x}{\sqrt N}\Big)^{\lambda-1}\Big]} dx\, ,  3<\lambda<5,
                \\
                 \int_{0}^{+\infty} \exp{\Big[-\frac{\langle k^2 \rangle x^2t}{2}-
                \frac{x^4}{N}\frac{\ln N}{24}\Big]}dx\, , \hspace{3em} \lambda=5,
                \\
                \int_{0}^{+\infty} \exp{\Big[-\frac{\langle k^2 \rangle x^2t}{2}-\frac{\langle k^4
                \rangle}{12}\frac{x^4}{N}\Big]}dx\, , \hspace{3em}   \lambda>5,
              \end{array}  \right.
\end{eqnarray}
with $t=(T-T_c)/T_c$ and the (pseudo)critical temperature
$T_c=\langle k^2 \rangle/\langle k \rangle$.\footnote{For a
finite-size system, $T_c$ depends on $N$ via the $N$-dependency of
the moments $\langle k \rangle$, $\langle k^2 \rangle$.} Similar to
 the previous section, the partition function  can be conveniently
represented in terms of a single variable $z$ that combines $t$ and
$N$ dependencies, cf. Eq. (\ref{II.13}). However, now this variable
differs in different regions of $\lambda$:
\begin{eqnarray}\label{III.19}
 z=\left\{
\begin{array}{ccc}
                & t \frac{\langle k^2 \rangle}{2}[a(\lambda)]^{2/(\lambda-1)}N^{\frac{\lambda-3}{\lambda-1}}\, , & 3<\lambda<5, \\
                 &t \langle k^2 \rangle\sqrt{6N}/\sqrt{\ln N}\ \, , & \lambda=5, \\
                &t \langle k^2 \rangle\sqrt{3N\langle k^4 \rangle}\, , & \lambda>5\, .
              \end{array}
  \right.
\end{eqnarray}
Written in terms of the variables appearing in (\ref{III.19}), the partition function (\ref{III.18}) has a simple form:
\begin{eqnarray}\label{III.20}
 Z(z)=\left\{
\begin{array}{ccc}
                & \int_{0}^{+\infty} \exp\Big(-zx^2-x^{\lambda-1}\Big)dx\, , & 3<\lambda<5, \\
                 &\int_{0}^{+\infty} \exp\Big(-zx^2-x^4\Big)dx \, , & \lambda\geq5 .
              \end{array}
  \right.
\end{eqnarray}
Two obvious conclusions follow: (i) since the functional form of
$Z(z)$ dependency  at $\lambda=5$ and $\lambda>5$ is the same, the
location of the Fisher zeros in the complex $z$   plane will be the
same too. This leads to the conclusion that an impact angle
$\varphi$ that corresponds to the Ising model on an annealed network
does not change for $\lambda\geq 5$ and (ii) these expressions
coincide with the partition function of the Ising model on the
complete graph, Eq. (\ref{II.12}). Therefore, analysis of  the
Fisher zeros of the last model, performed in section \ref{IIa}
applies equally well to the Fisher zeros of the  Ising model on an
annealed network. In particular, one concludes, that:
\begin{equation}\label{III.21}
\varphi=\pi/4,\hspace{2em} \lambda\geq 5\, .
\end{equation}
Let us note as well, that the logarithmic corrections to scaling
appear in the marginal case $\lambda=5$
\cite{Palchykov10,Igloi02,Loenedorog04,Kenna13}. Such a logarithmic
correction  appears in $I^\pm_\lambda(x)$ at $\lambda=5$ too.
However, it does not contribute to the terms leading in $1/N$,
resulting in the conclusion, that the impact angle of Fisher zeros
(and, therefore, the leading exponent for the heat capacity) is the
same at $\lambda=5$ and $\lambda>5$ .

\begin{figure}[t]
\centerline{
\includegraphics[angle=0, width=8.5cm]{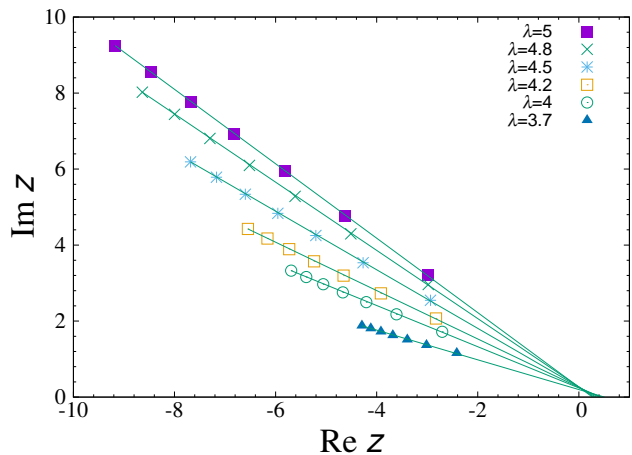}}
\caption{Fisher zeros for the Ising model on an annealed scale-free
network at zero magnetic field  for different  $\lambda$. The zeros
have a clear tendency to situate along the straight lines crossing
the real $z$   axis in the vicinity of the critical point $z_c$. The
angles formed by each of the lines with the real $z$ axis decrease
in the region $3<\lambda<5$ as predicted by equation (\ref{III.12}).
\label{fig11}}
\end{figure}

\begin{figure}[t]
\centerline{\includegraphics[angle=0, width=8cm]{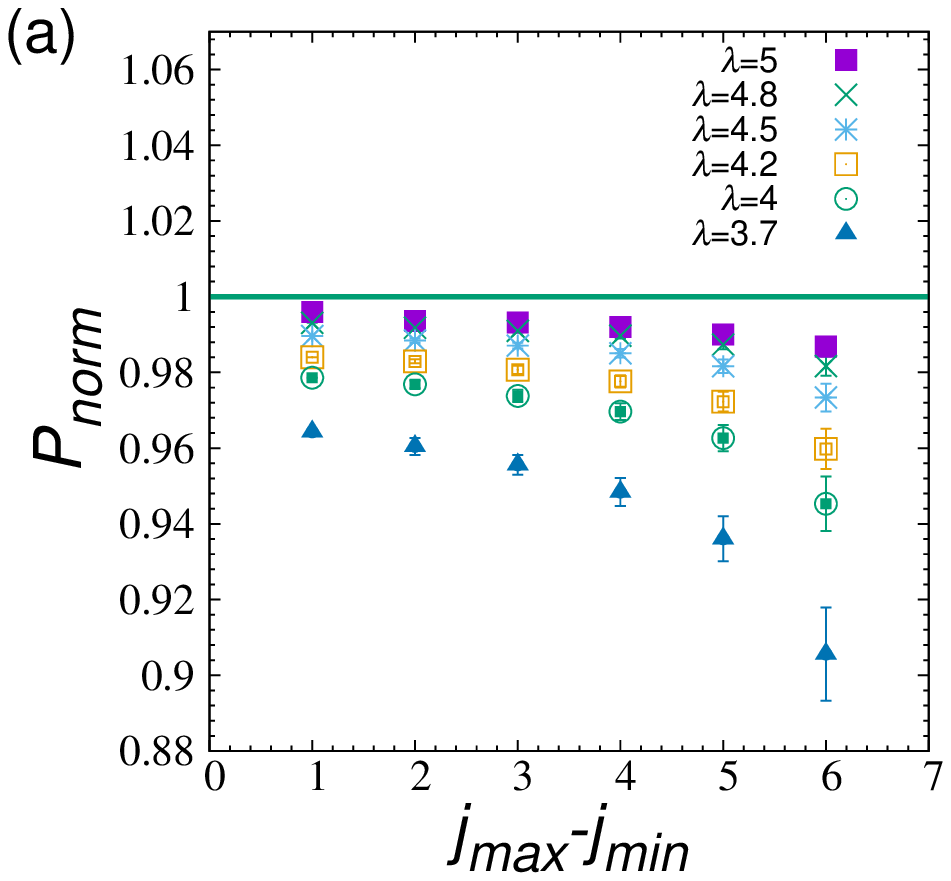}
\includegraphics[angle=0, width=8cm]{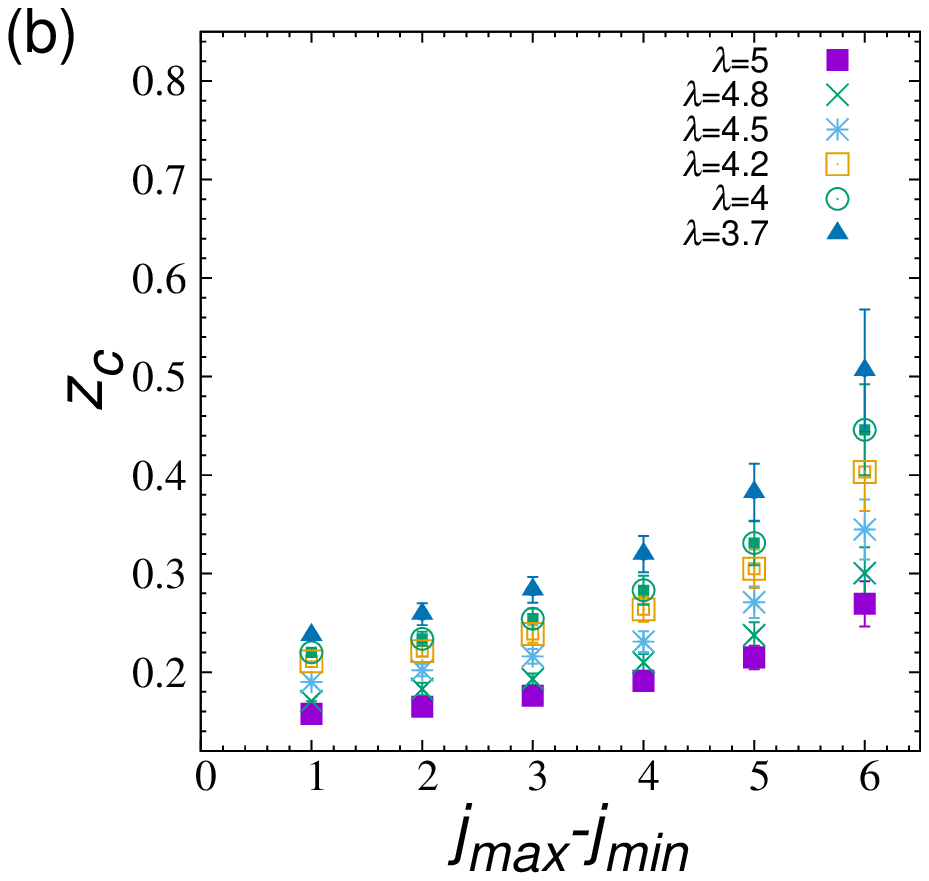}} \caption{Values (a) of the
ratio $P_{{\rm norm}}=\varphi/\frac{\pi(\lambda-3)}{2(\lambda-1)}$
and ({b}) of estimates for the critical temperature $z_c$ for the
partition function (\ref{III.20}) obtained by fitting of $j$ Fisher
zeros in the interval $j=j_{\rm min},\dots ,j_{\rm max}$ for $j_{\rm
max}=7$ and different values of $j_{\rm min}$. A solid line
in Panel {a} shows an exact value $P_{{\rm norm}}=1$. \label{fig12}}
\end{figure}

Similar to the preceding sections for the Ising model on the
complete graph, to proceed with the analysis of the Fisher zeros at
$3<\lambda<5$, we calculate the numerical values of the coordinates
of several first zeros  as shown in Fig.~\ref{fig11}. As
 is seen from the figure, the zeros have a clear tendency to
situate along the straight lines crossing the real $z$   axis in the
vicinity of the critical point $z_c$. Moreover, the angle formed by
each of the lines differs for different  $\lambda$ and decreases
with $\lambda$. To study this dependency in more detail, for each
value of $\lambda$ we fit $j$ Fisher zeros in the interval $j=j_{\rm
min},\dots ,j_{\rm max}$ for $j_{\rm max}=7$ and give the resulting
estimate for $\varphi$ in Table \ref{tab3}. One can see that the
values determined numerically approach those predicted by an
analytic formula (\ref{III.12}), the higher the order of the zeros
used for the fit, the higher the accuracy. This tendency is quite
similar to those observed for the Fisher zeros on a complete graph
(cf. Fig.~\ref{fig6}). To further demonstrate this resemblance, we
plot in Fig.~\ref{fig12}{a} the dependency of the ratio of the
numerically calculated angle $\varphi$ to its value predicted by
formula (\ref{III.12}): $P_{{\rm
norm}}=\varphi/\frac{\pi(\lambda-3)}{2(\lambda-1)}$. This ratio
tends to $P_{{\rm norm}}=1$ with an increase of the order of the
zeros used for the fit.  One can notice a similar tendency for the
behaviour of the critical temperature $z_c$, see
Fig.~\ref{fig12}{b}. Again one observes similarity with the
behaviour of $z_c$ for the Ising model on a complete graph, cf.
Fig.~\ref{fig6}{b}.

\begin{table}[b]
\caption{Numerically calculated values of the angle $\varphi$ for
different $\lambda$. The angle is calculated by linear fitting of
Fisher zeros with the indices $j=j_{\rm min},...,7$. The confidence
interval, when not written explicitly, is less than the last
significant digit. The last row gives $\varphi$ predicted by the
analytic formula (\ref{III.12}).
 \label{tab3}}
\begin{center}
\begin{tabular}{|c|c|c|c|c|c|c|c|c|}
  \hline
   $j_{\rm min}$& $\lambda\geq5$ & $\lambda=4.8$ & $\lambda=4.5$ & $\lambda=4.2$ & $\lambda=4$ & $\lambda=3.7$  \\
  \hline
  $1$ & $0.246(5)\pi$ & $0.233(1)\pi$ & $0.209(1)\pi$ & $0.180(1)\pi$ & $0.158(1)\pi$ & $0.117(2)\pi$ \\   \hline
  $2$ & $0.248\pi$ & $0.234\pi$ & $0.210\pi$ & $0.182\pi$ & $0.160(1)\pi$ & $0.121(1)\pi$ \\   \hline
  $3$ & $0.248\pi$ & $0.234\pi$ & $0.211\pi$ & $0.183\pi$ & $0.162\pi$ & $0.123\pi$ \\   \hline
  $4$ & $0.248\pi$ & $0.235\pi$ & $0.212\pi$ & $0.184\pi$ & $0.162\pi$ & $0.124\pi$ \\   \hline
  $5$ & $0.248\pi$ & $0.235\pi$ & $0.212\pi$ & $0.184\pi$ & $0.163\pi$ & $0.125\pi$ \\   \hline
  $6$ & $0.249\pi$ & $0.235\pi$ & $0.212\pi$ & $0.185\pi$ & $0.163\pi$ & $0.125\pi$ \\   \hline
   exact  & $0.250\pi$ & $0.237\pi$ & $0.214\pi$ & $0.188\pi$ & $0.167\pi$ & $0.130\pi$ \\   \hline
     \end{tabular}
\end{center}
\end{table}

Another inherent feature of the locations of the Fisher zeros  is
that the distance between two successive zeros decreases with
increasing index (see Fig.~\ref{fig11} for the  annealed scale-free
network as well as Fig.~\ref{fig5} for the complete graph). Indeed,
taken that the finite size scaling of the $j$-th zero $t_j$ in the
complex $t$   plane for large $j$ is given by \cite{Itzykson83}:
\begin{equation}\label{III.22}
t_j\sim\Big( \frac{j}{N} \Big)^\frac{1}{2-\alpha}
\end{equation}
one arrives at the conclusion that $\Delta t_j \equiv t_j - t_{j-1}
\sim j^{\kappa}$, with $\kappa=(\alpha-1)/(2-\alpha)$. The last
exponent is negative for the values of $\alpha$ we are interested
in: $\kappa=-1/2$, $\lambda\geq 5$ and $\kappa=-2/(\lambda-1)$,
$3<\lambda<5$. Rewriting (\ref{III.22}) in terms of $z_j$
(\ref{III.19}) gives:
\begin{equation}\label{III.23}
z_j\sim j^\frac{1}{2-\alpha}.
\end{equation}
To check how does the scaling of the Fisher zeros hold with $j$, we
calculate the value of the exponent in (\ref{III.23}) fitting the
function $\ln|z_j| = a + b\ln j$ for the Fisher zeros with indices
$j=j_{\rm min},...,7$ at different $\lambda$. The results are
compared with the exact value in Table \ref{tab4}. One can see the
right tendency of approach of the numerically calculated numbers to
their exact counterparts with an increase of $j$.
\begin{table}[b]
\caption{Linear fitting of the function $\ln|z_j| = a + b\ln j$ for
the Fisher zeros with indices $j=j_{\rm min},...,7$ at different
$\lambda$. The table shows results for the angle coefficient $b$.
The confidence interval, when not written explicitly, is less than
the last significant digit. The last row gives the value
$b=1/(2-\alpha)$ as predicted by an analytic formula (\ref{III.23}).
 \label{tab4}}
\begin{center}
\begin{tabular}{|c|c|c|c|c|c|c|c|c|}
  \hline
   $j_{\rm min}$& $\lambda>5$ & $\lambda=4.8$ & $\lambda=4.5$ & $\lambda=4.2$ & $\lambda=4$ & $\lambda=3.7$  \\
  \hline
  $1$ & $0.558(8)$ & $0.528(7)$ & $0.477(6)$ & $0.417(5)$ & $0.370(5)$ & $0.286(3)$ \\   \hline
  $2$ & $0.537(4)$ & $0.509(3)$ & $0.460(3)$ & $0.402(3)$ & $0.357(2)$ & $0.277(2)$ \\   \hline
  $3$ & $0.529(2)$ & $0.501(2)$ & $0.453(2)$ & $0.396(2)$ & $0.352(1)$ & $0.274(1)$ \\   \hline
  $4$ & $0.525(1)$ & $0.497(1)$ & $0.450(1)$ & $0.393(1)$ & $0.349(1)$ & $0.271(1)$ \\   \hline
  $5$ & $0.522(1)$ & $0.494(1)$ & $0.447(1)$ & $0.391(1)$ & $0.348(1)$ & $0.270(1)$ \\   \hline
  $6$ & $0.520$    & $0.492$    & $0.445$    & $0.390$ & $0.346$ & $0.269$ \\   \hline
  ${\rm exact}$& $0.500$& $0.474$& $0.429$ & $0.375$ & $0.333$ & $0.259$ \\   \hline
     \end{tabular}
\end{center}
\end{table}

In order to check scaling of the Fisher zeros $t_j$ with $N$, there
is no need to calculate explicitly their coordinates at different
values of $N$. Indeed, the $t_j(N)$ dependency follows from the
$z(t,N)$ functional form, as given by (\ref{III.19}). Expressing $t$
from there one gets that the power law scaling $t_j\sim
N^{1/(\alpha-2)}$ holds for $3<\lambda<5$ and $\lambda>5$ whereas it
is enhanced by a logarithmic correction at $\lambda=5$: $t_j\sim
N^{-1/2}(\ln N)^{1/2}$.

\subsection{Lee-Yang zeros for an annealed scale-free network at $T=T_c$ and motion of Fisher zeros in the magnetic field}\label{IIIc}

Substituting expressions for $I^\pm_\lambda(x)$
(\ref{III.15})--(\ref{III.17}) at $T=T_c=\langle k^2 \rangle/\langle
k \rangle$ into the partition function (\ref{III.9}) at different
values of $\lambda$ and keeping the leading contributions in $1/N$
we get:
\begin{eqnarray}\label{III.24}
 Z(h)=\left\{
\begin{array}{ccc}
                & \int_{0}^{+\infty} \exp\Big(-x^{\lambda-1}\Big)\cosh(hx)dx\, , & 3<\lambda<5, \\
                 &\int_{0}^{+\infty} \exp\Big(-x^4\Big)\cosh(hx)dx \, , & \lambda\geq 5\, .
              \end{array}
  \right.
\end{eqnarray}
Here, the $H$-    and $N$-dependencies of the partition function are
adsorbed in a single variable $h$. Its explicit form differs for
different regions of $\lambda$:
\begin{eqnarray}\label{III.25}
 h=\left\{
\begin{array}{ccc}
                & H \frac{\langle k \rangle^2}{\langle k^2 \rangle}a(\lambda)^{1/(1-\lambda)}N^{\frac{\lambda-2}{\lambda-1}}\, , & 3<\lambda<5, \\
                 &H \frac{\langle k \rangle^2}{\langle k^2 \rangle}\Big(\frac{24}{\ln N}\Big)^{1/4}N^{3/4}\ \, , & \lambda=5, \\
                &H \frac{\langle k \rangle^2}{\langle k^2 \rangle}\Big(\frac{12}{\langle k^4 \rangle}\Big)^{1/4}N^{3/4}\, , & \lambda>5\, .
              \end{array}
  \right.
\end{eqnarray}

Similar to the discussion for the $t_j(N)$ scaling at the end of
the previous subsection, here, the scaling of $h_j(N)$ directly
follows from the $h(H,N)$ dependency given by (\ref{III.25}).
Comparing (\ref{III.25}) with the definition (\ref{II.19}) one gets:
\begin{eqnarray}\label{III.26}
 \sigma=\left\{
\begin{array}{ccc}
                & \frac{\lambda-2}{\lambda-1}\, , & 3<\lambda<5, \\
                 & 3/4 \, , & \lambda >5\, .
              \end{array}
  \right.
\end{eqnarray}
The logarithmic correction appears at $\lambda=5$: $H_j \sim
N^{-3/4} \Big(\ln N\Big)^{1/4}$.

\begin{figure}[t]
\centerline{\includegraphics[angle=0,
width=8cm]{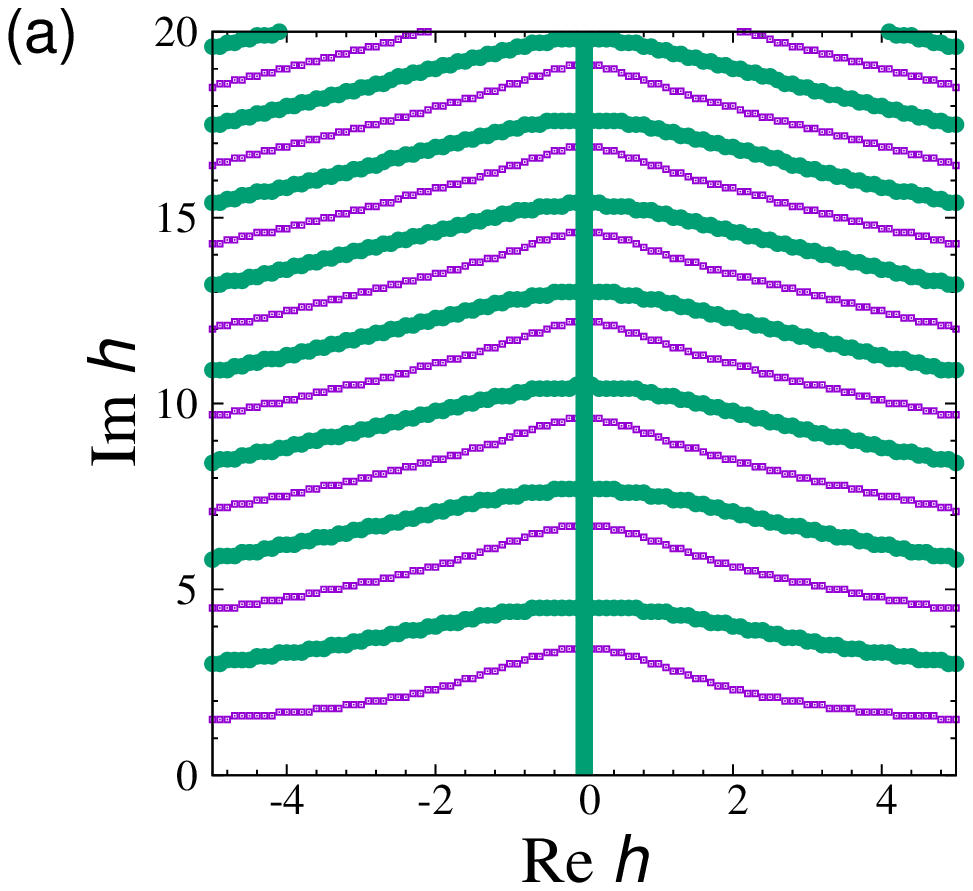}\includegraphics[angle=0,
width=8cm]{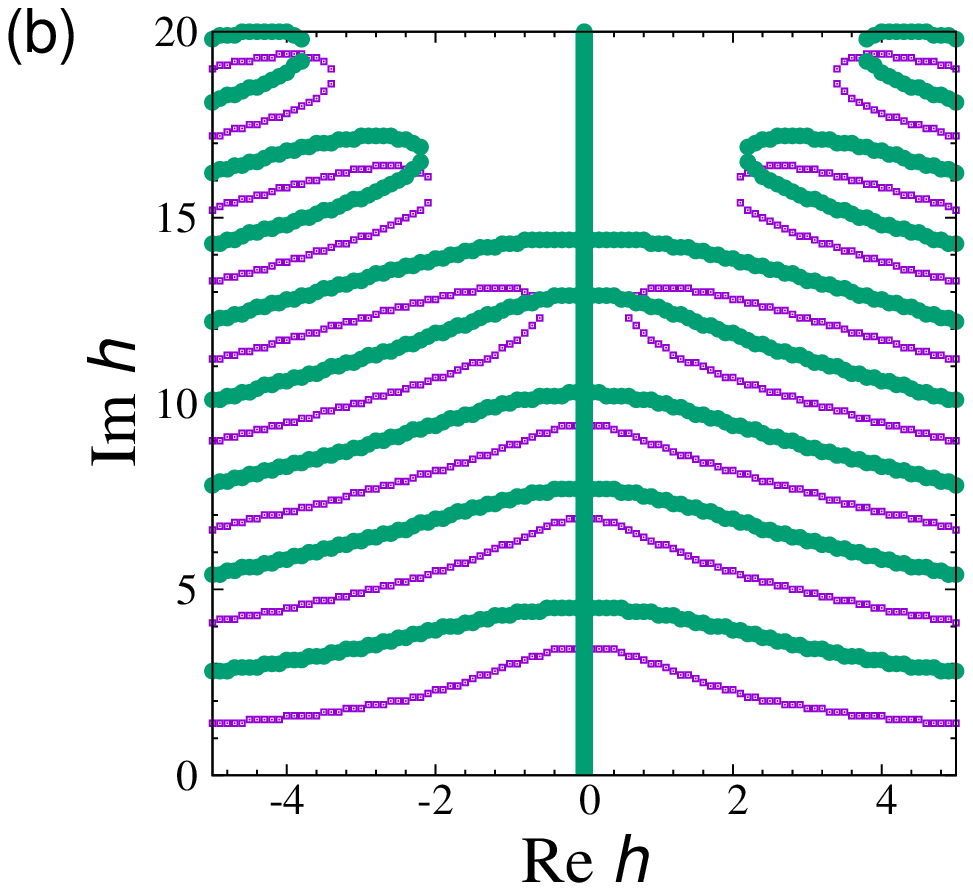}}  \vspace*{1pt}
 \centerline{\includegraphics[angle=0,
width=8cm]{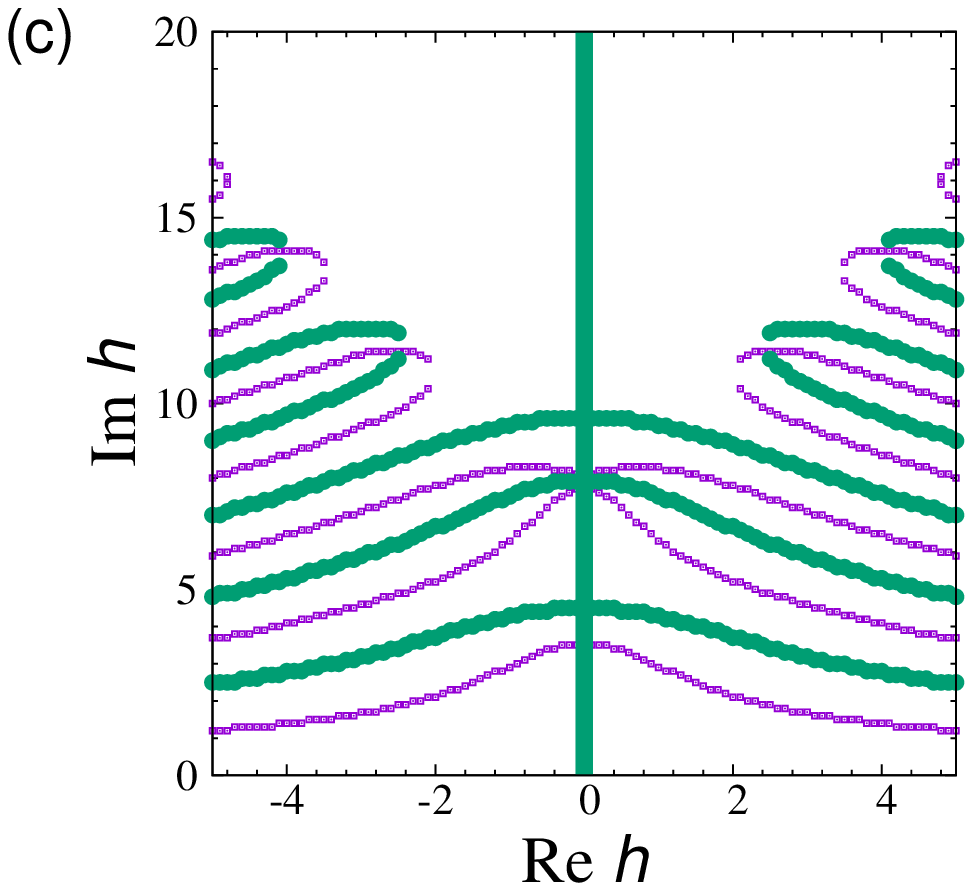}\includegraphics[angle=0,
width=8cm]{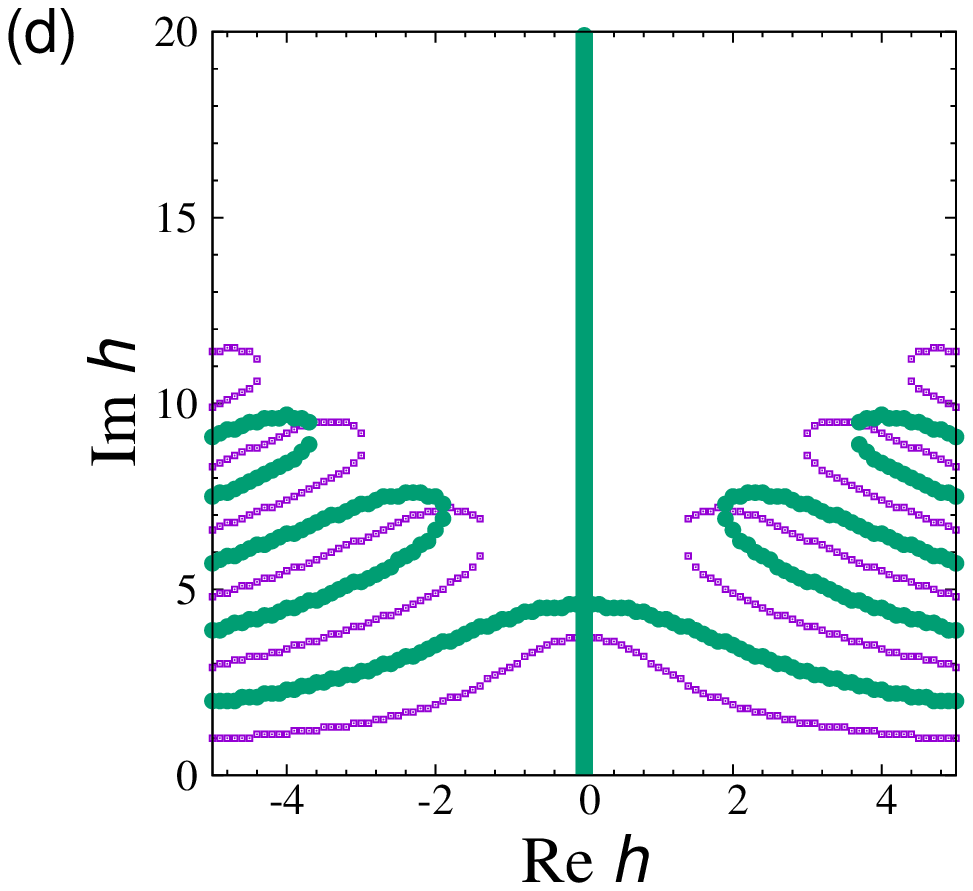}}
 \vspace*{1pt}
\caption{Lines of zeros for the real and imaginary part of the
partition function (\ref{III.24}) at $T=T_c$ and different values of
$\lambda$ in the complex magnetic field plane, { thin and thick
curves (violet and green dots online) respectively}: ({a})
$\lambda\geq5$; ({b}) $\lambda=4.5$; ({c}) $\lambda=4$; ({d})
$\lambda=3.5$. The points where the lines of different colours cross
give the coordinates of the Lee-Yang zeros. Note that one of the
${\rm Im}\, Z = 0$ lines coincides with the vertical axis in the
plot. \label{fig13}}
\end{figure}

In Fig.~\ref{fig13} we plot lines of zeros for the real and
imaginary parts of the partition function (\ref{III.24}) at $T=T_c$
and different values of $\lambda$ in the complex magnetic field
$h={\rm Re}\, h +i\,{\rm Im}\, h$ plane. The points where the lines
of different colour cross give the coordinates of the Lee-Yang
zeros. Note that whereas the coordinate of the first  Lee-Yang zero
(that closest to the origin) is purely imaginary for any value of
$3<\lambda<5$ (and it is its presence that allows to make
conclusions about the scaling with $N$ governed by the exponent
(\ref{III.26})), it is not the case for the zeros of higher order:
the number of zeros with ${\rm Re}\, h_j =0$ decreases with $j$. In
this respect the behaviour of Lee-Yang zeros for the Ising model on
an annealed scale-free network  differs crucially from the behaviour
on the complete graph (see the previous section). Whereas in the
complete graph the zeros were entirely imaginary and obeyed the
Lee-Yang circle theorem, this theorem is violated in the case of the
annealed scale-free network.

As  seen in Fig.~\ref{fig13}, the imaginary part of $Z(h)$ vanishes
when $h$ itself is imaginary. This is because the partition function
is an even function of $h$.
   The intersections of the different types of contour on the plot give the locations
of the Lee-Yang zeros. When $\lambda \geq 5$  all zeros  are on the
imaginary axis. But for $\lambda=4.5$ only first three zeros are
imaginary and zeros of higher order ($h_j$ for $j>3$) have
non-vanishing real parts. Similar behaviour is found for all values
of $\lambda$ between $3$ and $5$; there is a
 number, $\mathcal{N}$, finite in size, such that $h_j$ is purely imaginary for $j\le
{\mathcal{N}}$ but has non-vanishing real part for $j>\mathcal{N}$.
The number $\mathcal{N}$ decreases with decreasing $\lambda$.

 The values of Lee-Yang zeros determined numerically are listed in Table \ref{tab5}
for different values of $3<\lambda < \lambda_{\rm{uc}} = 5$, and for
$\lambda\geq 5$.  When $\lambda=4$, $\mathcal{N}=3$ while for
$\lambda = 3.5$, $\mathcal{N}=1$. In this respect the behaviour of
the zeros differs for $\lambda\geq5$ (where it is the same as for
the partition function of the Ising model on a complete graph) and
for $3<\lambda <5$. In Appendix~B we confirm this observation
showing that the asymptotic behaviour of the integral (\ref{III.24})
for non-integer $3<\lambda<5$ qualitatively differs from that at
$\lambda\geq 5$.

\begin{table}[b]
\caption{Coordinates of the few first Lee-Yang zeros $h_j$ for the
Ising model on an annealed scale-free network for different
$\lambda$. In the region $3<\lambda<5$ the number of purely imaginary zeros
 decreases with a decrease of $\lambda$. However,
the coordinate of the first Lee-Yang zero remains purely imaginary
for any value of $\lambda$. \label{tab5}}
\begin{center}
\begin{tabular}{|c|c|c|c|c|}
  \hline
  $j \backslash \lambda$ &  $\lambda>5$ & $\lambda=4.5$ &  $\lambda=4$    &  $\lambda=3.5$ \\
  \hline
  $j=1$& $i3.453$ & $i3.495$  & $i3.569$ &  $i3.762$  \\
  \hline
 $j=2$  &$i6.784$ & $i6.933$  &  $i7.823$ & $1.875+i7.212$   \\
  \hline
 $j=3$ & $i9.636$ &  $i9.474$ & $i
 8.149$  & $3.659+i9.496$      \\
  \hline
  $j=4$  & $i12.229$ &  $0.589+i12.848$   & $2.418+i11.466$  &  $5.138+i11.351$      \\
\hline
  $j=5$  & $i14.650$ &  $2.297+i16.346$  & $4.014+i14.174$   &  $6.435+i12.983$      \\
  \hline

   $j=6$  & $i16.945$ &  $3.761+i19.405$  & $5.446+i16.574$   &  $7.608+i14.470$      \\
  \hline
   $j=7$  & $i19.140$ &  $5.130+i22.229$  & $6.767+i18.776$   &  $8.690+i15.850$      \\
  \hline
   $j=8$  & $i21.254$ &  $6.427+i24.886$  & $8.005+i20.835$   &  $9.702+i17.146$      \\
  \hline
   $j=9$  & $i23.301$ &  $7.667+i27.414$  & $9.176+i22.783$   &  $10.656+i18.375$      \\
  \hline
   $j=10$  & $i25.289$ &  $8.859+i29.838$  & $10.293+i24.642$   &  $11.563+i19.548$      \\
  \hline
\end{tabular}
\end{center}
\end{table}

Since it has been established for the Ising model on a regular
lattice \cite{LeeYang52}, the Lee-Yang theorem has been
extended to a wider class of regular lattice Ising models.
 Besides the classical lattice discrete spin models \cite{Dunlop75}, the theorem also holds for
continuous spin systems \cite{Simon73}; for quantum systems such as
an ideal pseudospin $-1/2$ Bose gas in an external field and
arbitrary external potential \cite{Wang07};  for nonequilibrium
systems \cite{Blythe02}, which narrate the collective phenomena and
biophysics processes \cite{Loan98,Chou99}. There are many
models for which the Lee-Yang theorem does not hold:
 Ising models with antiferromagnetic interactions
 \cite{Kim04,Kim05}; with degenerated spins \cite{Suzuki73,Yamada81}; models with
multi-spin interactions at sufficiently high temperature
\cite{Monroe91,Chin87}; and van der Waals gases
\cite{Nilsen67,Hemmer64}. The theorem also fails for the model for
which the first order phase transition is observed: a high-$q$ Potts
model and the Blume-Capel model\cite{Lee94,Biskup00}. Theorem is
violated for the certain quantum many-body systems: Ising
ferromagnet with arbitrary spin \cite{Wang98} and the quantum
isotropic Ising chain in a transverse field \cite{Tong06}. For a
more detailed review see \cite{Bena05}.

\begin{figure}[t]
\centerline{
\includegraphics[angle=0, width=8.5cm]{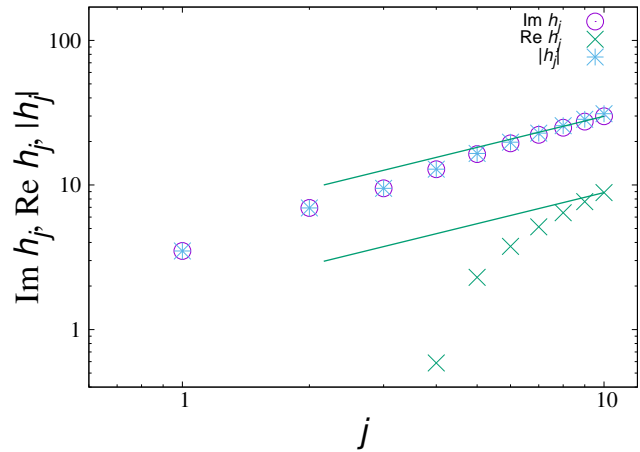}}
 \caption{ Coordinates of the first ten
Lee-Yang zeros $h_j$ of the Ising model on an annealed scale-free
network at $\lambda=4.5$. Different curves correspond to ${\rm Im}\,
h_j$, ${\rm Re}\, h_j$, $|h_j|$ as shown in the legend.  One can see
that the curves for ${\rm Im}\, h_j$ and $| h_j|$ are very close to
each other (since the value of ${\rm Re}\, h_j$ is relatively
small). The lines correspond to the expected value
$\sigma(\lambda=4.5)\simeq 0.714$, from (\ref{III.26}).
 \label{fig14}}
\end{figure}

Consequently, the Ising model on an annealed scale-free network at
$3<\lambda<5$ belongs to the group of models where the Lee-Yang
theorem is violated, whereas the model on the complete graph obeys
the theorem. The difference in behaviour of the Lee-Yang zeros of
the Ising model on a complete graph and on an annealed scale-free
network manifests itself also when one checks the scaling of the
zero coordinates with $j$. The coordinate of the zeros being
complex, the scaling in principle can be observed with respect to
the real or imaginary part of the coordinates or  their combination.
Typical results of our calculations are demonstrated in
Fig.~\ref{fig14}, where we plot coordinates of ${\rm Im}\, h_j$,
${\rm Re}\, h_j$ and $| h_j|$ for the first ten Lee-Yang zeros of
the Ising model on an annealed scale-free network at fixed value of
$\lambda=4.5$. The solid lines correspond to the expected value
$\sigma(\lambda=4.5)\simeq 0.714$, cf. (\ref{III.26}). One can see
that the curves for ${\rm Im}\, h_j$ and $| h_j|$  are very close to
each other, since the value of ${\rm Re}\, h_j$ is relatively small.
Similar behaviour is observed for the other values of $\lambda$. In
particular, for high $j$ the ${\rm Im}\, h_j$ and $| h_j|$ functions
are governed by the power law asymptotics with an exponent given by
(\ref{III.26}). {{The power law behaviour is less pronounced for
small $j$. Introducing a more general form of scaling that involves
the parameter $C$ as it was done for the complete graph, see
equation (\ref{II.19}), does not help to improve the picture.}}

Let us next consider  the motion of the Fisher zeros in the real
magnetic field.  To this end, similar as it was done in the former
section for the Ising model on the complete graph, we present the
partition function of the Ising model on an annealed scale-free
network  as a function of rescaled variables $z$ and $h$.
Substituting expansions (\ref{III.15})--(\ref{III.17}) for the
functions $I^\pm_\lambda(x)$ into the partition function
(\ref{III.9}) one arrives at the following representations:
\begin{eqnarray}\label{III.27}
 Z(t,h)=\left\{
\begin{array}{ccc}
                & \int_{0}^{+\infty} \exp\Big(-tx^2-x^{\lambda-1}\Big)\cosh(hx)dx\, , & 3<\lambda<5, \\
                 &\int_{0}^{+\infty} \exp\Big(-tx^2-x^4\Big)\cosh(hx)dx \, , & \lambda\geq
                 5\, ,
              \end{array}
  \right.
\end{eqnarray}
where $t$ and $h$ are defined by the Eqs. (\ref{III.19}) and
(\ref{III.25}) respectively.

\begin{figure}[t]
\centerline{
\includegraphics[angle=0,width=8cm]{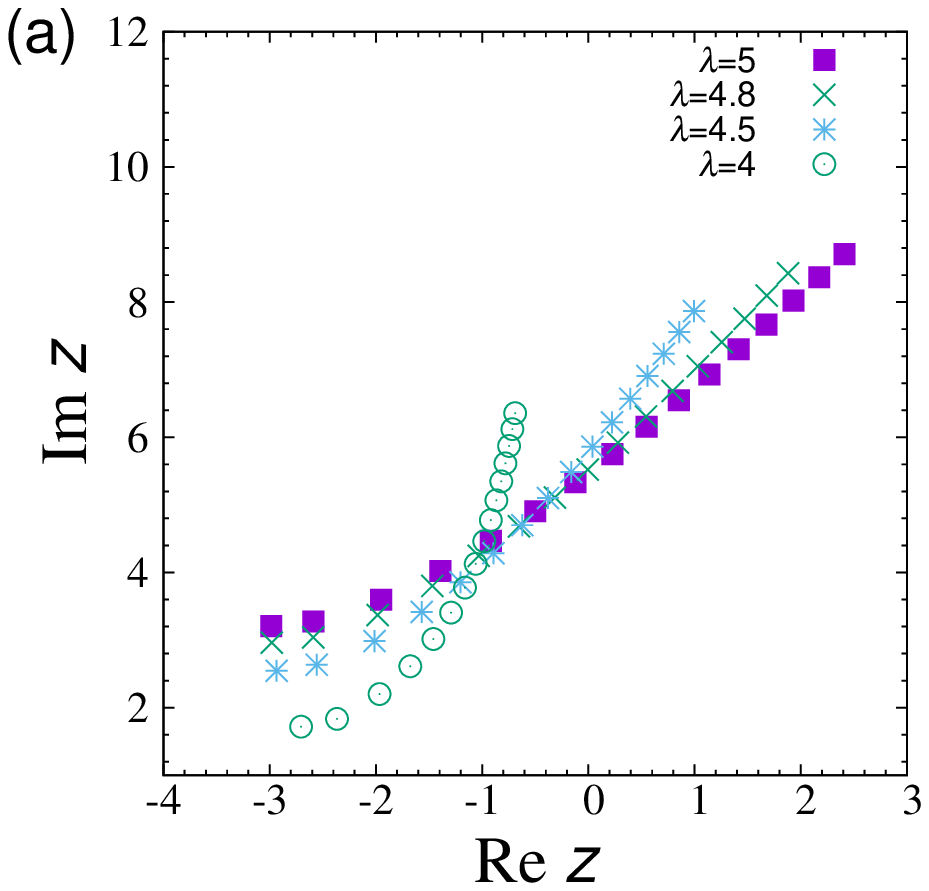}
\includegraphics[angle=0,
width=8cm]{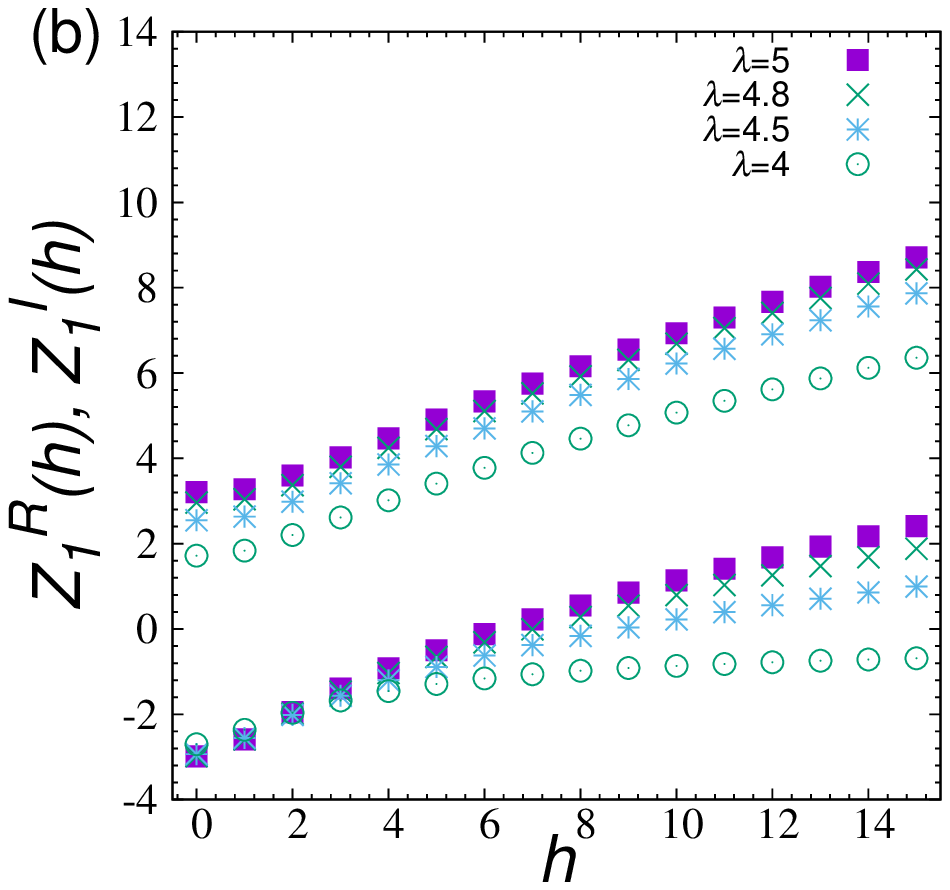}}
 \vspace*{1pt}
\caption{({a}) Motion of the first Fisher zeros of the partition
function (\ref{III.27}) in the complex $z$ plane for different
values of the real magnetic field $h=s$, $s=0,1,\dots,15$ and
$\lambda$. ({b}) Scaling functions ${\cal Z_{\it 1}^{\,R}}\, (h)$
(lower plots),
 ${\cal Z_{\it 1}^{\,I}}\, (h)$
 (upper plots) (\ref{III.28}) for the real and imaginary part of the first Fisher
 zero coordinate $z_1$, as functions of the scaling variable
$h$ at different values of $\lambda$. \label{fig15}}
\end{figure}

Fig.~\ref{fig15}{a} shows results obtained for the motion of the
Fisher zeros in the real magnetic field calculated at different
$\lambda$ for functions (\ref{III.27}). The squares correspond to
$\lambda\geq5$ and the expected value of the angle of motion from
(\ref{III.13}) is $\psi=\pi/3$. Therefore, in this region of
$\lambda$ our numerical results are in a good agreement with the
analytical prediction as we get $\psi\simeq 59^\circ $. However,
this is not the case for the smaller values of $3<\lambda<5$.
Indeed, naively substituting critical exponents values into the
equation (\ref{I.12}) for the angle $\psi$ one gets
 \begin{equation}\label{III.13}
\psi=\frac{\pi(\lambda-3)}{2(\lambda-2)}\, .
 \end{equation}
In particular, from (\ref{III.13}) one expects that the angle of
motion will decrease with $\lambda$. The three curves from the
Fig.\ref{fig15}{a}, calculated for $\lambda=4.8, \, 4.5$ and 4
demonstrate the opposite behaviour: the angle increases with
decreasing $\lambda$ with a tendency to reach asymptotics at
$\psi=\pi/2$. As we have shown above, for the model under
consideration the zeros  at the critical point do not obey the
Lee-Yang theorem: there is only a  small number of zeros with purely
imaginary coordinates for $3<\lambda<5$. Therefore the equation
(\ref{III.13}) that was obtained under an assumption that this
theorem holds  is not applicable. The angle $\psi$ (\ref{I.12})
describing the motion of Fisher zeros in real magnetic field does
not seem to be related by a simple scaling relation to the critical
exponents.

Similar to what was done for the Ising model on a complete graph,
one can define scaling functions for the real and imaginary parts of
the  $j$th zero $z_j$. Using the definition (\ref{II.34B}) of the
scaling functions from the subsection \ref{IIc} for ${\cal Z_{\it
j}^{\,R}}\, (h)$, ${\cal Z_{\it j}^{\,I}}\, (h)$ in different
regions of $\lambda$ we define the scaling functions for the Ising
model on an annealed scale-free network as
\begin{eqnarray}\label{III.28}
{\cal Z_{\it j}^{\,R,\, I}}\, (h) \equiv\left\{
\begin{array}{lll}
& \frac{\langle k^2 \rangle}{2}[a(\lambda)]^{2/(\lambda-1)}\, {\cal
T_{\it
j}^{\,R,\, I}}\,\Big(h\frac{\langle k^2 \rangle}{\langle k \rangle^2}a(\lambda)^{1/(\lambda-1)}\Big)\ \, , & 3<\lambda<5, \\
&\langle k^2 \rangle\sqrt{6}/\sqrt{\ln N} {\cal T_{\it
j}^{\,R,\, I}}\, \Big(h\frac{\langle k^2 \rangle}{\langle k \rangle^2}(\frac{\ln N}{24})^{1/4}\Big)\ \, , & \lambda=5, \\
&\langle k^2 \rangle\sqrt{3\langle k^4 \rangle}{\cal T_{\it
j}^{\,R,\, I}}\, \Big(h\frac{\langle k^2 \rangle}{\langle k
\rangle^2}(\frac{\langle k^4 \rangle}{12})^{1/4}\Big)\, , &
\lambda>5\, ,
\end{array}
\right.
\end{eqnarray}
and expressions for $h$ at different $\lambda$ are given by
(\ref{III.25}). The scaling functions for the first Fisher zero
 $z_1$ at different values of $\lambda$ are plotted in
Fig.~\ref{fig15}{b}. From Eq. (\ref{II.35}) we expect that the ratio
${\cal Z_{\it 1}^{\,I}}\, (0)/{\cal Z_{\it 1}^{\,R}}\, (0)$ gives
the Fisher pinching angle $\varphi$, which in the case of the
scale-free network is $\lambda$-dependent (see subsection
\ref{IIIb}). From  Fig.~\ref{fig15}{b} we calculate ${\cal Z_{\it
i}^{\,I}}\, (0)/{\cal Z_{\it i}^{\,R}}\, (0)=0.921;\, 0.798;\,
0.577$ or $\varphi \simeq 0,249\pi;\, 0.228\pi;\, 0.18\pi$ for
$\lambda=4,8;\, 4.5;\, 4$ respectively. These values agree with
those predicted by equation (\ref{III.12}) and given in Table
\ref{tab3}. Note that the agreement increases with an increase of
the zero number $j$.

\section{Conclusions and outlook}\label{IV}

In this paper we analyse the properties of the complex zeros of the
partition function for the case of the Ising model on graphs of two
types. First, we considered the case of the complete graph and then,
we addressed the new features that are introduced by the changes in
graph topology, by considering the Ising model on an annealed
scale-free network. While the former case have been already a
subject of several studies (see e.g. \cite{Glasser87,Uzelac13}), the
latter is addressed here for the first time.

In particular, using the approach of Ref.
\cite{Glasser86} we have analyzed the behaviour of Lee-Yang zeros as
well as tracked the motion of the Fisher zeros in the real external
magnetic field.
As  has been observed before for some  other
models \cite{Itzykson83,Zuber83}, this motion is governed by the
universal angle $\psi$, that encodes values of the order parameter
critical exponents through (\ref{I.12}). A fundamental feature of
the Ising model on the complete graph, as well as of those models
where such behaviour of Fisher zeros in real magnetic field was
observed, is that their zeros in complex magnetic field at the critical
temperature  are governed by the Lee-Yang theorem:  plotted in
the plane $e^{-h}$ they align along the unit circle \cite{LeeYang52}.

Analysis of section \ref{III} has shown that complex zeros behaviour
of the Ising model on an annealed scale-free network is
characterised by several new and rather unexpected features. First
of all, as far as the universality class of the model is defined by
the node-degree distribution decay exponent $\lambda$ (\ref{III.8}),
the pinching angle $\varphi$ as well as the Lee-Yang edge exponent
$\sigma$ are $\lambda$-dependent too. Their dependence on $\lambda$
in the region $3<\lambda<5$ is given by equations (\ref{III.12}),
(\ref{III.14}). In turn, for the fast-decaying degree distributions
with $\lambda>5$ we recover the values obtained previously for the
Ising model on the complete graph. Secondly, the logarithmic
corrections to leading scaling behaviour that arise for the spin
models on scale-free networks
\cite{Igloi02,Loenedorog04,Palchykov10,Kenna13} at $\lambda=5$ are
observed for the exponents that govern condensation of Lee-Yang and
Fisher zeros with an increase of the system size. Corresponding gaps
in $H$- and $t$   planes decrease with $N$ at $\lambda=5$ as
\begin{equation}\label{IV.1}
H_j\sim N^{-3/4} \Big(\ln N\Big)^{1/4}\, ,\hspace{2em} t_j\sim
N^{-1/2}(\ln N)^{1/2}\, .
\end{equation}
Note, that the powers of the logarithms comply with the
corresponding scaling relations \cite{Kenna13}.

The striking feature of the Ising model on an annealed scale-free
network is that its partition function zeros calculated at $T_c$ in
complex magnetic field does not obey the Lee-Yang theorem. The
number of purely imaginary zeros decreases with the decrease of
$\lambda$, and the zeros acquire both real and imaginary parts. Note
that experimental identification of the Lee-Yang
zeros \cite{Peng15} has so far  only be made for purely imaginary zeros
 \cite{Wei12}. The results identified herein show that
while these are accessible for lattices and complex networks for
sufficiently large $\lambda$, not all network zeros are accessible
in this manner for $\lambda < 5$. A challenge for experiment is to
find another way to access them.

As we have already mentioned, there are some other models, for which
the Lee-Yang theorem does not hold
\cite{Kim04,Kim05,Suzuki73,Yamada81,Monroe91,Chin87,Nilsen67,%
Hemmer64,Lee94,Biskup00,Wang98,Tong06}.
Unlike the above examples, the problem we have considered in this
paper concerns ferromagnetic Ising model, and the Lee-Yang theorem
was proven \cite{Lieb81} to hold  for any Ising-like model with
ferromagnetic interaction, see also \cite{Kozitsky}. In this respect
the analyzed here Hamiltonian (\ref{III.1}) looks as if it will lead
to the Lee-Yang property of the partition function too, since the
adjacency matrix elements $J_{lm}$ are non-negative. Furthermore, in
the course of partition function calculation, in the spirit of the
mean-field approximation, the matrix elements $J_{lm}$ are
substituted by $k_lk_m>0$, see Eq. (\ref{III.4}). Again, this should
not spoil the Lee-Yang property for any fixed choice of the node
degree $k_l$. Next step in the derivation was the integration over
$k$ with the distribution function $p(k)$, Eq. (\ref{III.6}). This is the place which leads to the violation of the Lee-Yang unit
circle theorem: indeed, if one takes a sum (or an integral) of
functions each of which possesses the Lee-Yang property, the sum
might not possess such property at all. That is exactly what is
observed for the annealed network.

\section*{Acknowledgements} This work was supported in part by FP7
EU IRSES projects  No. $295302$ ``Statistical Physics in Diverse
Realizations", No. $612707$ ``Dynamics of and in Complex Systems",
No. $612669$ ``Structure and Evolution of Complex Systems with
Applications in Physics and Life Sciences", and by the Doctoral
College for the Statistical Physics of Complex Systems,
Leipzig-Lorraine-Lviv-Coventry $({\mathbb L}^4)$. It is our pleasure
to thank Yuri Kozitsky, Taras Krokhmalskii, Lo\"ic Turban, and
Jean-Yves Fortin for useful comments and discussions. M.K. is
grateful to Taras Hvozd for the help with programming.

\section*{APPENDIX A. Notes on the natural cut-off}
In this Appendix we discuss in more detail the choice of the upper
cut-off  in Eq. (\ref{III.7}).
A network of finite size
(finite number of nodes $N$) cannot contain a node of infinite
degree $k$ by definition.
For a scale-free network, the maximal node
degree $k_{\rm max}$ for finite $N$ (the so-called natural cut-off)
can be defined in different ways.
 In Ref. \cite{Aiello01} it was
proposed to define a cut-off $k_{\rm max}$ as the value of the
degree for which one expects to observe one vertex at most:
\begin{equation}\label{AA1}
Np(k_{\rm max}) \sim 1.
\end{equation}
From here one gets an $N$ dependency of the cut-off:
\begin{equation}\label{AA2}
k_{\rm max} \sim N^{1/\lambda}.
\end{equation}
The drawback of (\ref{AA2}) is that for continuous $k$ this
definition contains the probability of a single point (which is
zero). Another definition of cut-off was given in
\cite{Dorogovtsev02}. There, the cut-off is interpreted as the value
of the degree $k$ above which one expects to find at most one
vertex:
\begin{equation}\label{AA3}
N\int_{k_{\rm max}}^\infty p(k) d k \sim 1,
\end{equation}
leading to
\begin{equation}\label{AA4}
k_{\rm max} \sim N^{1/(\lambda-1)}.
\end{equation}
In Ref. \cite{Boguna04} the extreme value theory was used to define
the natural cut-off. In this case the exact value for $k_{\rm max}$
slightly differs, however it remains $N$ dependent. Below we will
show that the form of $k_{\rm max}(N)$ dependence does not change the
asymptotic estimate for the partition function (\ref{III.9}) in the
leading order in $N$.

For simplicity, let us consider the partition function (\ref{III.5})
for zero magnetic field $H=0$ (the derivation for non-zero $H$ is
quite similar):
\begin{equation}\label{AAIII.1}
Z_N (T)= \int_{-\infty}^{+\infty} \exp\Big (\frac{-N\langle k
\rangle x^2T}{2}+\sum_{l}\ln \cosh( xk_l)\Big)dx \, .
\end{equation}
Using (\ref{III.6}) we rewrite the sum over network nodes $\sum_{l}$
in the exponent in terms of the integral over $k$ for a given
distribution function $p(k)$, keeping the lower and upper cut-off:
\begin{eqnarray}\label{AAIII.5}
Z_N (T)&=& \\ \nonumber  \int_{-\infty}^{+\infty}  && \exp\Big \{N
\Big [\frac{-\langle k^2 \rangle x^2\, t }{2}+ \int_{k_{\rm
min}}^{k_{\rm max}}p(k)[\ln \cosh( xk) - \frac{1}{2} ( xk)^2]
dk\Big] \Big \} dx \, ,
\end{eqnarray}
with $t=(T-T_c)/T_c$ and $T_c= \langle k^2\rangle/\langle k\rangle$.
To evaluate (\ref{AAIII.5}) we make use of the fact that at large
$N$ the main contribution  comes from small $x$.
In turn, this means that we need to evaluate the integral over $k$
in (\ref{AAIII.5}) at small $x$ and for large $N$. To keep the
network connected for all $3<\lambda<5$ we have chosen the lower
cut-off value to be $k_{\rm min}=2$ (see section
\ref{IIIa}). The crucial point is that the value of this integral
does not depend on the way in which $k_{\rm max}$ tends to infinity (e.g.
$k_{\rm max}\sim \ N^{\frac{1}{\lambda-1}}$). It is because $k_{\rm
min}$ and $k_{\rm max}$ are the only terms in (\ref{AAIII.5}) that
depend on $N$. Moreover, this dependence may be taken in the form of
the
 natural cut-off (\ref{AA4})  or in any other form that satisfies
 $\lim_{N\to\infty}k_{\rm max}\to \infty$. Therefore, one gets:
\begin{eqnarray}\label{AAIII.6}
\lim_{x\to 0} \lim_{N \to \infty} \int_{k_{\rm min}}^{k_{\rm
max}}p(k)[\ln \cosh( xk) - \frac{1}{2} ( xk)^2] dk =
\\ \nonumber
\lim_{x\to 0}
\int_{2}^{\infty}p(k)[\ln \cosh( xk) - \frac{1}{2} ( xk)^2] dk \, .
\end{eqnarray}
The integral in the right-hand side of (\ref{AAIII.6}) has been
already evaluated (see e.g. \cite{Igloi02} or \cite{Krasnytska13}).
It is
\begin{equation}\label{AAIII.7}
\lim_{x\to 0}  \int_{2}^{\infty}p(k)[\ln \cosh( xk) - \frac{1}{2} (
xk)^2] dk = a_\lambda x^{\lambda-1} + O(x^\lambda),
 \end{equation}
with known values of the coefficients $a_\lambda$
\cite{Igloi02,Krasnytska13}. Substituting (\ref{AAIII.7}) into
(\ref{AAIII.5}) we arrive at:
\begin{equation}\label{AAIII.8}
Z_N(\tau) \sim \int_{0}^{\infty}\exp\Big \{-N\Big (\frac{\langle k^2
\rangle x^2 \tau}{2}+ a(\lambda)x^{\lambda-1}\Big)\Big \}dx\, ,
\hspace{2em} 3<\lambda<5\, .
\end{equation}
Now the integral (\ref{AAIII.8}) can be taken by a steepest descent.
The result we have obtained is consistent with the Landau theory for
the Ising model on scale-free networks \cite{Goltsev03} as well as
with the mean field analysis \cite{Igloi02,Loenedorog04}. In this
way, all three approaches (Landau, mean field, and the current one -
based on the integral representation) lead to the same
thermodynamics.

\section*{APPENDIX B. Asymptotic estimates for the integral (\ref{III.24}) at
non-integer $\lambda$}

In this Appendix we determine an asymptotic estimate for the
behaviour of the partition function $Z(h)$ (\ref{III.24}) at ${\rm
Re}\, h=0$ in the limit of large ${\rm Im}\, h$. To this end, let us
consider an integral
\begin{equation}\label{A1}
Z(ir)=\int_{0}^{\infty} e^{-x^{\lambda-1}}\, \cos(rx)dx\, ,
\end{equation}
that depends on the real variable $r$. At $r={\rm Im}\, h$
(\ref{A1}) gives an imaginary part of the partition function
(\ref{III.24}). Asymptotic behaviour of $Z(ir)$ follows from the
Erdelyi lemma (see e.g. \cite{Fedoruk87}) as explained below.

The Erdelyi lemma gives an asymptotic behaviour of the integral:
\begin{equation}\label{A2}
F(y)= \int_0^A x^{b-1}f(x)e^{iyx^a}dx.
\end{equation}
According to the lemma, if $a\geq1$, $b>0$ and the function $f(x)$
becomes zero together with all its derivatives at the upper
integration limit: $f(A)=f'(x)|_{x=A}=...=f^{(n)}(x)|_{x=A}=0$, the
following asymptotic estimate for the integral (\ref{A2}) is valid:
\begin{equation}\label{A3}
F(y)\sim \sum_{k=0}^\infty a_k y^{-\frac{k+b}{a}}\, , \hspace{1cm}
y\rightarrow\infty \, ,
\end{equation}
where the coefficients $a_k$ are given by
\begin{equation}\label{A4}
a_k=\frac{f^{k}(0)}{k!a}\Gamma\Big (\frac{k+b}{a}\Big ) \exp\Big
(\frac{i\pi(k+b)}{2a} \Big)\, .
\end{equation}
Changing variables in (\ref{A1}) one can represent $J_\lambda(y)$ in
the form similar to (\ref{A3}):
\begin{equation}\label{A5}
Z(ir) = \frac{5}{\lambda-1}{\rm Re}\, \int_0^\infty
x^{\frac{5}{{\lambda-1}}-1}e^{-x^5}e^{irx^{5/({\lambda-1})}}dx\, .
\end{equation}
Note that the upper integration limit is this case is $a=\infty$. It
is easy to see that the conditions of the lemma are satisfied and an
asymptotic expansion for the function (\ref{A1}) follows:
\begin{equation}\label{A6}
Z(ir)\sim \sum_{k=0}^\infty b_k r^{-\frac{k (\lambda-1)}{5}-1}\,
,\hspace{1cm} r\rightarrow\infty \,,
\end{equation}
where the coefficients $b_k$ are defined as
\begin{equation}\label{A7}
b_k=\frac{(\lambda-1)f^{k}(0)}{5k!}\Gamma\Big
(\frac{k(\lambda-1)}{5}+1 \Big ) \cos \Big
(\frac{\pi(k(\lambda-1)+5)}{10} \Big )\, .
\end{equation}

\begin{figure}[t]
\centerline{\includegraphics[angle=0, width=8cm]{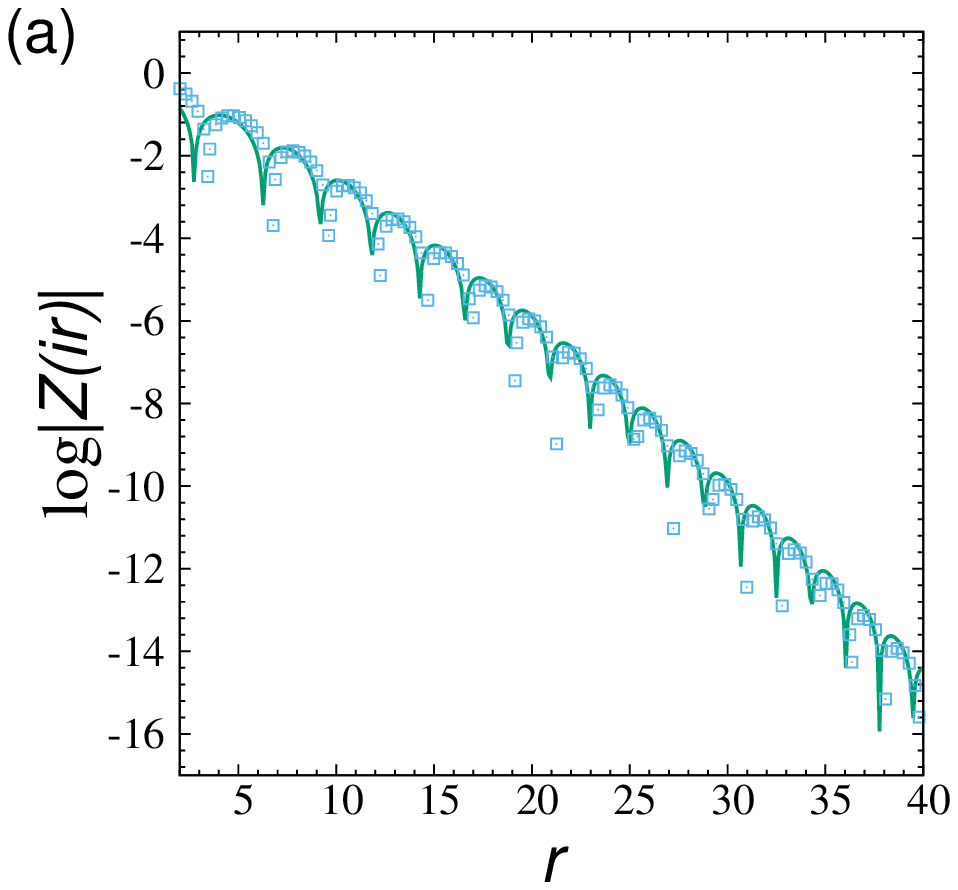}
\includegraphics[angle=0, width=8cm]{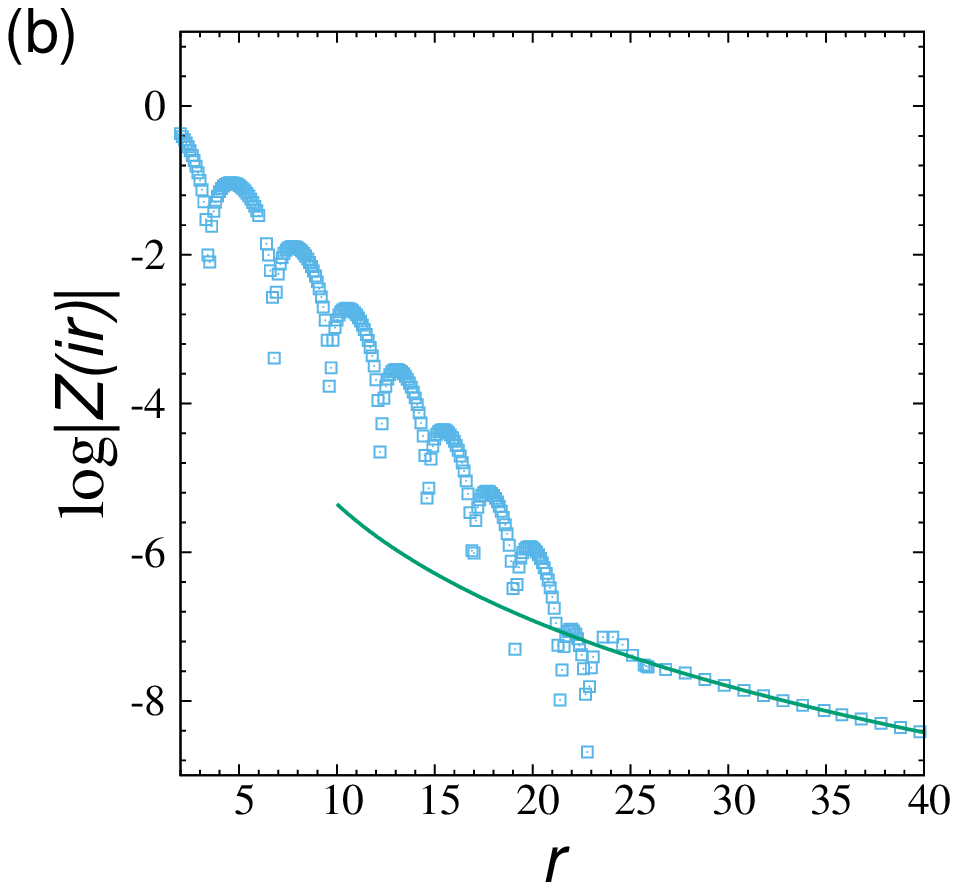}}  \vspace*{1pt}
 \centerline{\includegraphics[angle=0, width=8cm]{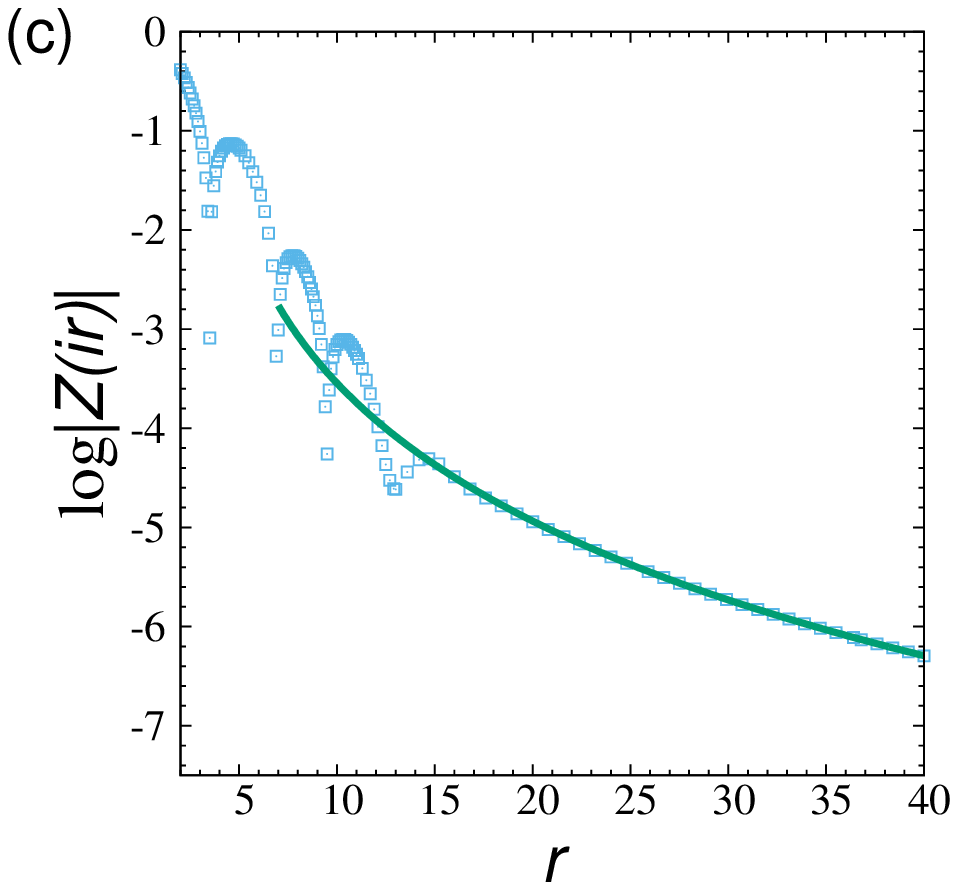}}
 \vspace*{1pt}
\caption{{ Function  $\log |Z (ir)|$ for: ({a}) $\lambda\geq5$,
({b}) $\lambda=4.99$, and ({c}) $\lambda=4.5$. Solid lines: results
of asymptotic expansions. Squares: numerically calculated values.
For $\lambda\geq 5$ the function keeps oscillating in the
asymptotics (meaning that  the number of zeros is unbounded), it is
not the case for $\lambda<5$.} \label{fig16}}
\end{figure}

Note, that coefficients $b_k=0$ for the marginal values $\lambda=3$
and $\lambda=5$: the decay differs from the power law. To evaluate
it, we proceed as follows.
\begin{itemize}
\item $\lambda=3$

For $\lambda=3$ the integral (\ref{A1}) is taken exactly leading to
\begin{equation}\label{A8}
Z(ir)=\sqrt {\pi/2} \exp\Big(-r^2/4\Big)\, .
\end{equation}
Therefore, in the limit of large $r$ the integral decays
exponentially. Moreover, $Z(ir)>0$ for any $0<r<\infty$ which
signals about an absence of partition function zeros on the complex
$h$   plane at $\lambda=3$. In turn, this brings about an absence of
the phase transition in this case.

\item $\lambda=5$

In this case integral (\ref{A1}) can be rewritten as:
\begin{equation}\label{A9}
Z(ir)=\int_{0}^{+\infty} e^{-x^{4}}\, \cos(rx)dx\,= \frac{1}{2} {\rm
Re}\, \int_{-\infty}^{+\infty} e^{-x^{4}}\, e^{ixr}  dx \, .
\end{equation}
Its asymptotics  can be evaluated by a steepest descent method,
calculating the function under the integral in the point of extremum
at $x=(ir/4)^{1/3}$ and leading to:
\begin{equation}\label{A10}
Z(ir)\sim \frac{1}{2}\exp\Big(-\frac{3}{2}(r/4)^{4/3}\Big ) \cos\Big
(\frac{3\sqrt{3}}{2}(r/4)^{4/3} \Big) \, ,\hspace{1cm}
r\rightarrow\infty \, .
\end{equation}
The integral $Z(ir)$ has an infinite number of zeros due to the
presence of an oscillating function in the r.h.s., as we further
demonstrate below.
\end{itemize}

In Fig.~\ref{fig16} we compare behaviour of the numerically
calculated function $\log |Z (ir)|$  with its asymptotic expansion
using Eqs. (\ref{A6}) and (\ref{A10}) for different $\lambda$. One
can see that behaviour of the function for $\lambda<5$ is
qualitatively different from that at $\lambda\geq 5$: whereas in the
last case the function keeps oscillating in the asymptotics (meaning
that  the number of zeros is unbounded), it is not the case for
$3<\lambda<5$. Here, after finite number of oscillations the
function approaches its asymptotics from above. Therefore, the
number of zeros is limited. As one can see from the
Figs.~\ref{fig16}{a}, {c}, {d}, the number of oscillation decreases
with the decrease of $\lambda$.

\section*{References.}

\providecommand{\newblock}{}

\end{document}